%% file: main.tex
\documentclass[sigconf]{acmart}
\renewcommand\footnotetextcopyrightpermission[1]{} 
\usepackage{booktabs} 

\usepackage{algorithmic}
\usepackage{graphicx}
\usepackage{textcomp}
\usepackage{xcolor}
\usepackage{blindtext}
\usepackage{afterpage}
\usepackage{tabularx}
\usepackage{colortbl}
\usepackage{hhline}
\usepackage{tabularx}

\usepackage{xspace}

\usepackage{amsmath,empheq}

\newcommand\floor[1]{\lfloor#1\rfloor}

\newtheorem{proposition}{\textbf{Proposition}}[section]
\newtheorem{definition}{\textbf{Definition}}[section]
\newcommand{\plcwm}{\emph{PLC Watermarking}\xspace}
\newcommand{\sct}{\emph{Scan Cycle Time}\xspace}

\newtheorem{thm}{\textbf{Theorem}}[section]

\usepackage{subfig}

\usepackage{array}
\usepackage{siunitx}
\usepackage{caption}
\usepackage{longtable}
\usepackage{graphicx}
\usepackage{subfiles}
\usepackage[export]{adjustbox}
\usepackage{multirow} 
\usepackage{booktabs} 
\usepackage{longtable} 
\usepackage{rotating} 
\usepackage{enumitem}

\acmConference[AsiaCCS'21]{ACM AsiaCCS conference}{June 2021}{Hong Kong}
\acmYear{2021}
\copyrightyear{2021}

\begin{document}
\title{Scanning the Cycle: Timing-based Authentication on PLCs}



\author{Chuadhry Mujeeb Ahmed}
\affiliation{%
  \institution{University of Strathclyde}
  \country{United Kingdom}}
  \email{mujeeb.ahmed@strath.ac.uk}

\author{Martin Ochoa}
\affiliation{%
  \institution{AppGate}
  \country{Colombia}}
\email{martinochoa@gmail.com}

\author{Jianying Zhou}
\affiliation{%
  \institution{Singapore University of Technology and Design}
  \country{Singapore}}
\email{jianying\_zhou@sutd.edu.sg}

\author{Aditya Mathur}
\affiliation{%
  \institution{Singapore University of Technology and Design}
  \country{Singapore}}
\email{aditya\_mathur@sutd.edu.sg}






\begin{abstract}
 
Programmable Logic Controllers~(PLCs)  are a core component of an Industrial Control System (ICS). However, if a PLC is compromised or the commands sent across a network from the PLCs are spoofed, consequences could be catastrophic. 
In this work, a novel technique to authenticate PLCs is proposed that aims at raising the bar against powerful attackers while being compatible with real-time systems. The proposed technique captures  timing information for each controller in a non-invasive manner. It is argued that  \emph{Scan Cycle} is a unique feature of a PLC that can be approximated passively by observing network traffic. An attacker that spoofs commands issued by the PLCs would deviate from such  fingerprints.
To detect replay attacks a \plcwm technique is proposed. \plcwm models the relation between the scan cycle and the control logic by modeling the input/output as a function of request/response messages of a PLC. The proposed technique is validated on an operational  water treatment plant (SWaT) and smart grid (EPIC) testbed. Results from experiments indicate that PLCs can be distinguished   based on their \emph{scan cycle timing characteristics}.
\end{abstract}

\maketitle 

\input{body.tex}



\bibliographystyle{ACM-Reference-Format}
\bibliography{main.bib}

\input{appendix.tex}

\end{document}

%% file: body.tex
\section{Introduction}
An Industrial Control System~(ICS) uses sensors to remotely measure the system state and then feeds the sensor measurements to a Programmable Logic Controller~(PLC). 
PLCs send the control actions to the actuators based on sensor measurements. PLCs also share local state measurements with other PLCs via a messaging protocol. 
ICSs are an attractive target for cyber attacks~\cite{ICS_CERT_2014} due to the critical nature of ICS infrastructures and therefore, demand security measures for safe operations~\cite{cardenas2009challenges}. Recent research efforts in ICS security stem from a traditional IT infrastructure perspective.
Network-based intrusion detection is a widely proposed solution~\cite{CPS_security_survey2017}. 
However, conventional network traffic based intrusion detection methods would fail when an attacker impersonates a PLC and there would be no change in network traffic patterns~\cite{can_fingerprinting_ecus_usenix2016}. 
Most commercial ICS communication protocols lack integrity checks, resulting in no data integrity guarantees~\cite{CPS_security_survey2017}. In several of such  protocols, no authentication measures are implemented, and  hence an attacker can manipulate data transmitted across the PLCs and devices, e.g., the actuators~\cite{fovino_2009_malware_PLC}. 

Although solutions grounded in cryptography, such as those that use TLS, HMACs or other authentication and/or 
integrity guarantees have been advocated in the 
context of ICS, historically such countermeasures are not widespread due to limitations in hardware and relative 
computational cost of such protocols ~\cite{John_ACNS2017}. Since many ICSs run legacy hardware, and are intended to do so for several years, the problem of raising the bar against authentication attacks by non-cryptographic means is a practical one. 
A recent study reported in~\cite{eireann_2013_greyhat_PLC_vulnerabiity} reveals that a large number of PLCs are connected to the Internet and contain vulnerabilities related to authentication. 
Also, the use of commercial off the shelf~(COTS) devices in an ICS, and software backdoor, can lead to full control over PLCs~\cite{ruben_backdoors_PLC_2012_blackhat}. 
Stuxnet is a famous example of a malware attack where PLCs were hijacked and malicious code altered the PLC's configuration~\cite{stuxnet_langner_2011_SnP}. 
A range of malware and network-based attacks were designed and executed against PLCs~\cite{Anand_ESORICS2017_PLC_ladderlogicbomb}. Therefore, there is a need for enhancing authentication in PLCs non-invasively and without disturbing their core functionality. We present two techniques in the following.


\noindent \textit{PLC Fingerprinting}: The PLC fingerprint is a function of its hardware and  control functionality, i.e., the timing characteristics of a PLC. Inspired by timing based fingerprinting in other domains, it is argued that there is a unique feature of PLCs known as \emph{scan cycle}. A \emph{scan cycle} refers to the periodic execution of the PLC logic and input/output (I/O) read/write. This unique feature of PLCs is being used here as its fingerprint. The challenge here is to create a fingerprint in a passive manner without disturbing the system's functionality. 
\emph{Scan cycle} timing is estimated in a non-invasive manner by monitoring the messages which are being exchanged between the PLCs. 
Uniqueness in the fingerprint is due to the hardware components such as clock~\cite{kohno2005}, processor, I/O registers~\cite{iFinger_register_fingerprint_JSAC2020}, and logic components, e.g., control logic, message queuing, etc. 
An adversary can send malicious messages either by using an external device connected to the ICS network, or as Man-in-The-Middle~(MiTM)~\cite{urbina_CCS2016limiting}, to modify the messages, or from outside of the system to perform DoS attacks~\cite{robert_turk_2005_PLC_DoS,Anand_ESORICS2017_PLC_ladderlogicbomb}. These attacks, even if launched by a knowledgeable attacker, would be detected since the timing profile resulting from the injected data would not match the reference pattern representative of the unique characteristics of a PLC. In general, it is shown that any attack on PLC messages could be detected if it changes the statistics of estimated \emph{scan cycle} timing distribution. 

\noindent \textit{PLC Watermarking}: In addition, this paper proposes a novel solution to detect advanced replay and masquerade attacks. 
The proposed technique is called \plcwm. \plcwm is built on top of \sct estimation and the dependency of such an estimate on the control logic. A \emph{PLC Watermark} is a random delay injected in the control logic and this watermark is reflected in the estimated \sct. This leads to the detection of powerful masquerade and replay attacks because PLC watermark behaves as a nonce. 
Experimental results on a real-world water treatment~(SWaT) testbed available for research~\cite{swat2016}, support the idea of fingerprinting the timing pattern for PLC identification and attack detection. Experiments are performed on a total of six Allen Bradley PLCs available in the SWaT testbed and four Wago PLCs, four Siemens IEDs in EPIC testbed~\cite{Nandha_ESORICS2020_EpicDataset}. Results demonstrate that PLC identification and attack detection can be performed with high accuracy. 
Moreover, it is also shown that although our methodology can raise false positives, the rate at which they are raised is practical, in the sense that it can be managed by a human operator without creating bottlenecks, or can be fed to metamodels that take into account other features (such as model-based countermeasures, IDS alarms, etc.).

\noindent \textbf{Contributions}: In summary, this paper proposes a novel non-cryptographic risk-based technique to authenticate  PLCs and detect attacks.  There have been several research works on network intrusion detection systems using network traffic features~\cite{sommer_paper_snp2010_ml_challenges}. However, it is known that anomaly detection in inter-arrival time of packets alone does not work well in practice \cite{sommer_paper_snp2010_ml_challenges}. 
 Our contributions are thus: 
\textbf{a)} A novel technique to fingerprint PLCs by exploiting \emph{scan cycle} timing information.
\textbf{b)} A novel \plcwm technique to detect a powerful cyber attacker that is aware of timing profiles used for fingerprinting, e.g., replay attacks.

 \begin{figure}[!htb]
\centering
\includegraphics[scale=.3]{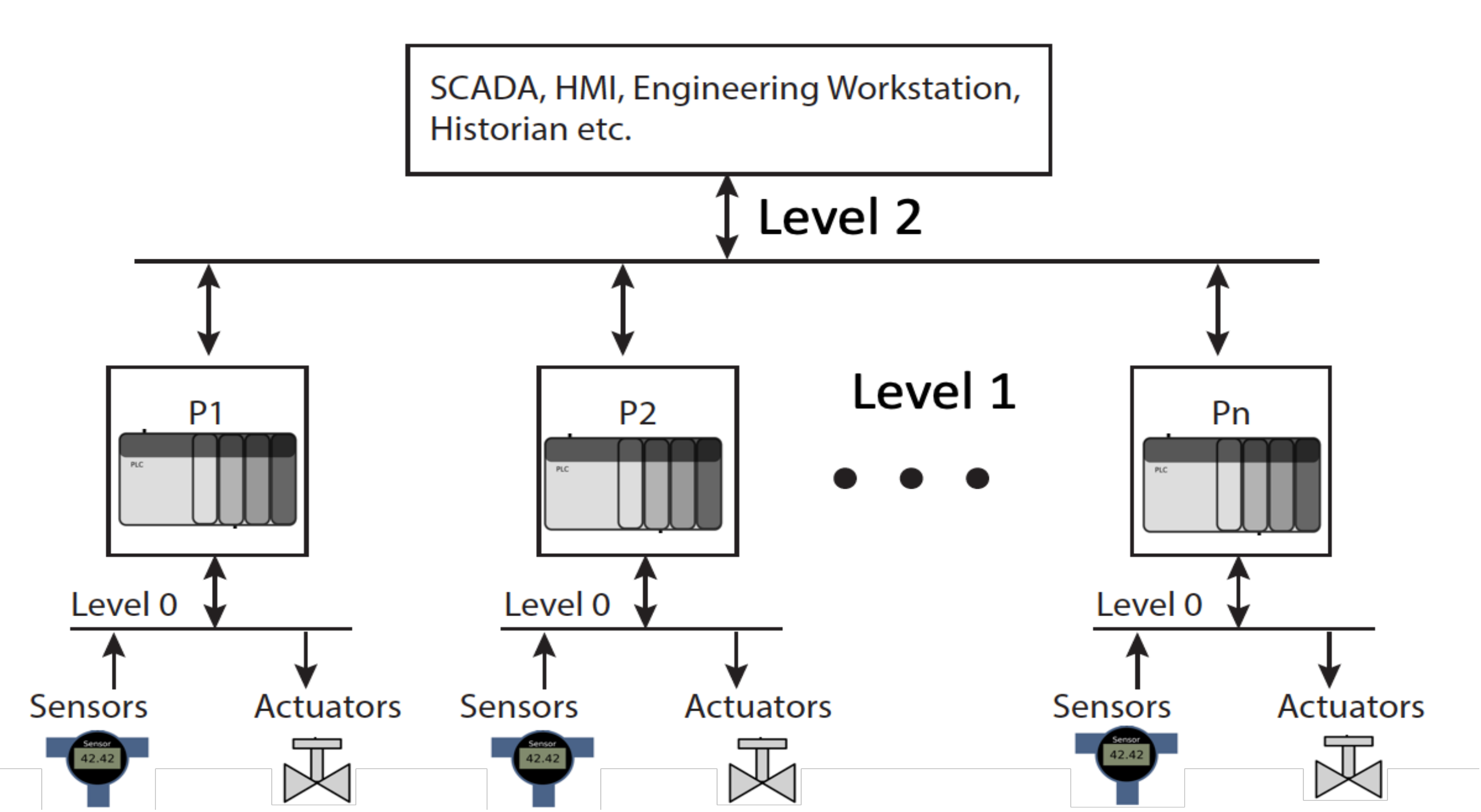}
\caption{Overview of a common ICS network architecture.}
\label{comm_arch_fig}
\vspace{-1.5em}
\end{figure}


\begin{figure}[t]
\centering
\includegraphics[height=5cm,width=12cm,keepaspectratio]{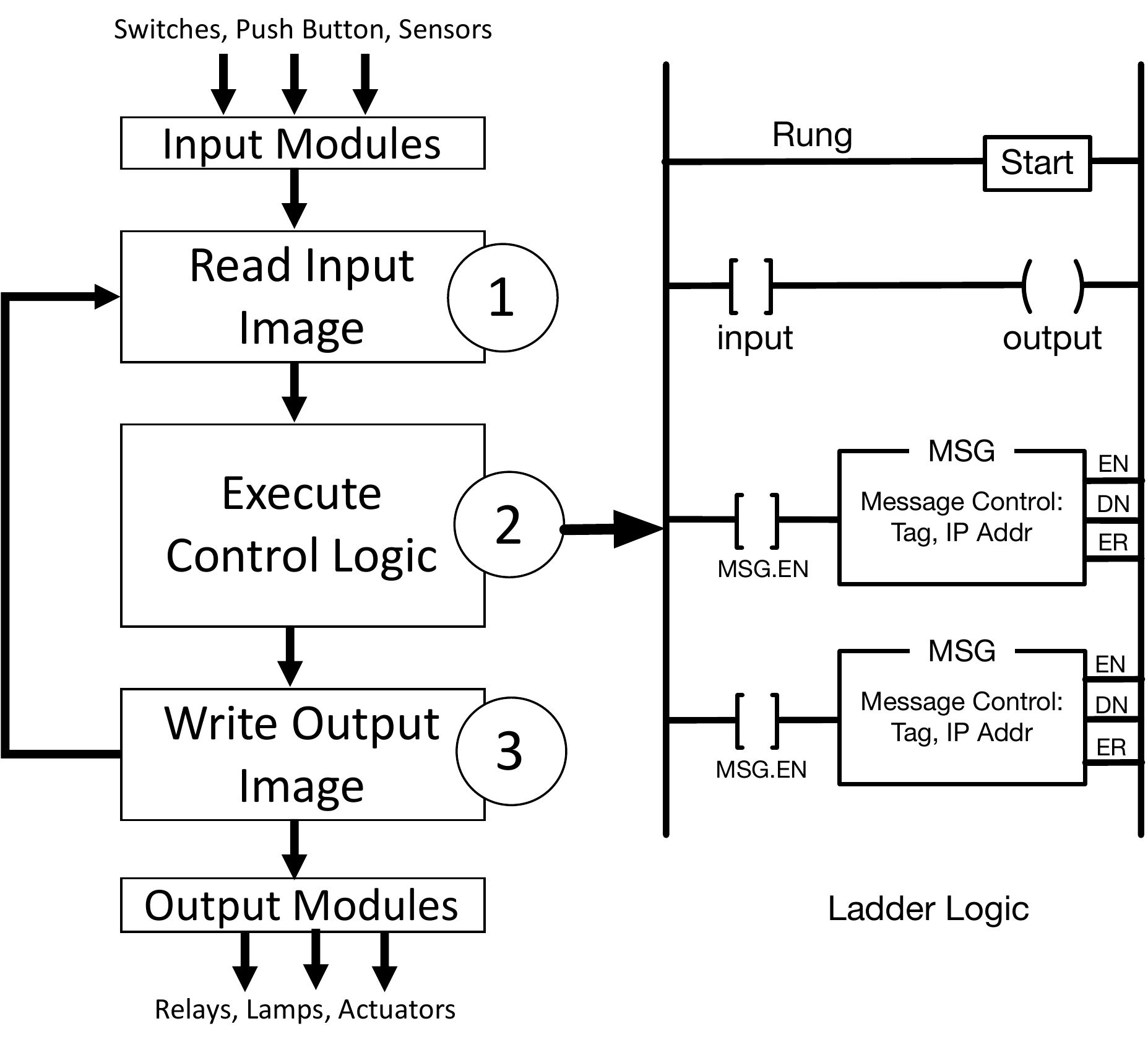}
\caption{A \emph{Scan Cycle} and ladder logic.}
\label{scancycle_ladderlogic_fig}
\vspace{-1.5em}
\end{figure}

\section{System and Attacker Model} 


\label{system_model_Sec}


\subsection{Architecture of an ICS}

A typical ICS is composed of field devices (e.g., sensors and actuators), control devices (e.g., PLCs), as well as SCADA, HMI and engineering workstations. In general an ICS follows a layered architecture~\cite{Williams_purdue_reference_architecture}. As shown in  Figure~\ref{comm_arch_fig}, there are three  levels of  communications. Level~0 is  the field communication network and is composed of field devices, e.g.,  remote I/O units and communication interfaces to send/receive information to/from PLCs. 
Level~1 is the communication layer where PLCs communicate with each other to exchange data to make control decisions. Level~2 is where PLCs communicate with the SCADA workstation, HMI, historian server; this  is the supervisory control network. The communication protocols in an ICS have been proprietary until recently when the focus shifted to using the enterprise network technologies for  ease of deployment and scalability, such as  the Ethernet and TCP/IP~\cite{communication_protocol_ICS_survey_2013}. 

\subsection{PLC Architecture and the Scan Cycle}

A PLC consists of a central unit called the processor, a program and data memory unit, input/output~(I/O) interfaces, communication interfaces and a power supply. I/O interface connects the PLC with input devices, e.g., sensors and switches and output devices, e.g, actuators. The communication interfaces are used to communicate with other devices on the network, e.g., a human-machine interface~(HMI), an engineering workstation, a programming device and other PLCs. 


\emph{Scan Cycle Time~($T_{SC}$)}: The PLCs are part of real-time embedded systems and have to perform time-critical operations. To optimize this objective, there is the concept of control loop execution in the PLCs. A PLC has to perform its operations continuously in a loop called the \emph{scan cycle}. There are three major steps in a \emph{scan cycle}, 1) reading the inputs, 2) executing the control logic, 3) writing the outputs. 
A \emph{scan cycle} is in the range of milliseconds~(ms) with a strict upper bound referred to as the \emph{watchdog} timer else PLC enters fault mode~\cite{allenbradley_manual_execution_procedure}. 
The duration of the \emph{scan cycle} time is based on several factors including the speed of the processor, the number of I/O devices, processor clock, and the complexity of the control logic. 
Therefore, with the variations in the hardware and control logic, these tasks take variable time even for the same type of machines, resulting in device fingerprints as shown previously for the personal computers~(PC)~\cite{gtid_raheem_device_fingerprinting_tdsc2015}. 
Figure~\ref{scancycle_ladderlogic_fig} shows the logical flow of the steps involved during a PLC \emph{scan cycle}. 
Expression for the \emph{scan cycle} time can be written as,

\begin{equation}
\label{scan_cycle_eq}
T_{SC} = T_{IN} + T_{CL} + T_{OP}.
\end{equation}

\noindent Where $T_{SC}$ is the \emph{scan cycle} time of a PLC, $T_{IN}$ is the input read time, $T_{CL}$ is the control logic execution time, and  $T_{OP}$ is the output write time. In this work, the challenge is to estimate the \emph{scan cycle} time~($T_{SC}$) in a non-invasive manner and create a hardware and software fingerprint based on the uniqueness of the \emph{scan cycle} in each PLC. To further understand the proposed technique, the relationship of the network communication to the \emph{scan cycle} is elaborated in the following section.  

\subsection{Monitoring the \emph{Scan Cycle} on the Network}
It is possible to capture the \emph{scan cycle} information using a system call in the PLC but that would not be useful to detect network layer attacks. The idea is to get the \emph{scan cycle} timing information outside the PLCs and in a passive manner. To this end, it is proposed to estimate the \emph{Scan Cycle} over the network and refer  to it as the \emph{Estimated Scan Cycle~($T_{ESC}$)} time. Communication between PLCs is based on a request-response model. 
 The message exchange between different PLCs can be programmed using the message instruction~(\emph{MSG}) on a ladder rung using the control logic as shown in Figure~\ref{scancycle_ladderlogic_fig}.
Since the \emph{MSG} instruction is executed in the control logic which is step~2 of the \emph{scan cycle}, this \emph{MSG} instruction would occur at this specific point in the \emph{scan cycle}. 
The \emph{scan cycle} time~($T_{SC}$) can be estimated  by observing the MSG requests being exchanged among PLCs on the network layer. 

\subsection{Scan Cycle vs. Time~(IAT) of MSG Instructions}

Are Scan Cycle and IAT (inter-arrival time) of MSG equivalent? The answer is \emph{No}. Although the idea of exploiting MSG instructions in the control logic to obtain \emph{scan cycle} information is intuitive, it turns out to be challenging. If the MSG instructions were executed each \emph{scan cycle} then by monitoring the IAT of MSG instructions alone would have provided  \emph{scan cycle} information. Based on this idea a measurement experiment is conducted, for which results are reported in Table~\ref{responseTime_table}. In Table~\ref{responseTime_table} $E[T_{SC}]$ represents the mean of the \emph{scan cycle} time measured inside a PLC and $E[T_{ESC}]$ represents the mean of the MSG instructions IAT. It turns out that MSG instructions IAT instead is equal to a multiple of the scan cycle time, referred to as estimated \emph{scan cycle} in this paper. Since  messages are analyzed at the network layer it is important to find out the relationship between the \emph{scan cycle} of a PLC and what is observed at the network layer. On the network, a message would be seen at the following intervals,

\begin{equation}
T_{ESC} = T_{Proc} + T_{Txn} + T_{Prog} + T_{SC} + T_{Que},
\end{equation}
where $T_{Proc}$ is the packet processing delay at a PLC, $T_{Txn}$ and $T_{Prog}$ are the packet transmission and propagation delays respectively. $T_{Que}$ represents the queuing delay. The relationship between $T_{SC}$ and $T_{ESC}$ can be simplified to:
\begin{equation}
T_{ESC} = T_{OverHead} + T_{SC}.
\end{equation}
Where $T_{OverHead} = T_{Proc} + T_{Txn} + T_{Prog} + T_{Que}$ is the time it takes for a packet to get processed at the PLC, enter a message queue for transmission and finally get transmitted on the network. From the previous research on the network delays, it is known that the transmission and propagation delays are fixed per route and does not influence the variation in delays of the packets~\cite{moon1999}, while variable queuing delay has a significant effect on the packet delay timings and its reception on the network~\cite{moon1999}. Since the network configuration and the number of connected devices for an ICS network are fixed, the propagation delays can be measured and treated as constant. The significant effect on the estimation of the \emph{scan cycle} is due to the queuing delay. The randomness in the queuing delay depends on the network traffic directed to a particular PLC and its processor usage. The next step is to quantify that delay and figure out if it still reveals the information about the \emph{scan cycle}. It is true that the message instructions are scanned in each \emph{scan cycle} but their execution depends on the two conditions~\cite{allenbradley_MSG_instruction_queuing} as specified in the following.



\begin{figure}[t]
\centering
\includegraphics[width=7cm]{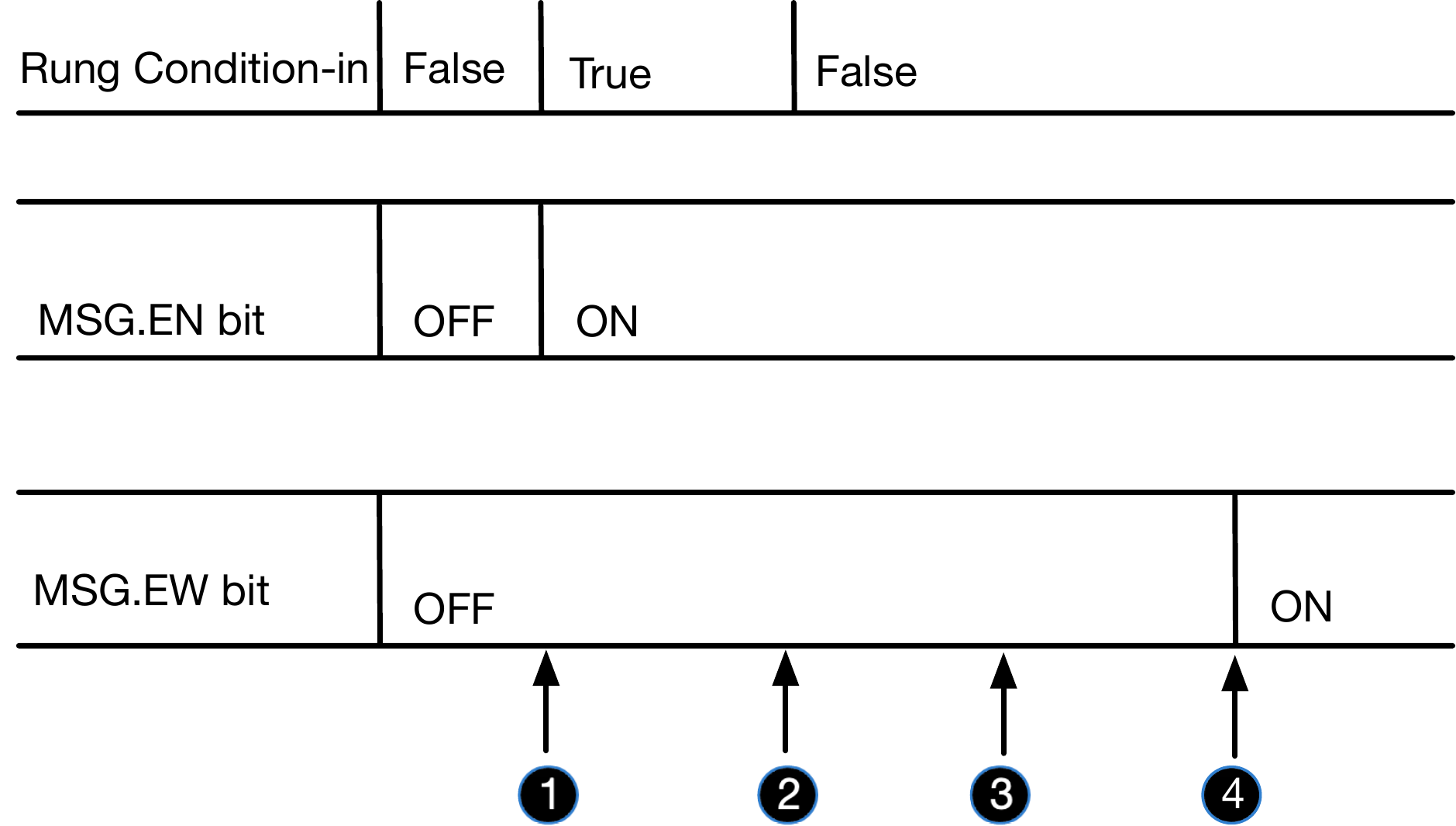}
\caption{Message Queue in AB PLCs~\cite{allenbradley_MSG_instruction_queuing}}
\label{msg_queue_fig}
\vspace{-1em}
\end{figure}

\vspace{0.2cm}
\fbox{\begin{minipage}{24em}
\underline{\emph{Condition 1:}} The response for the previous message request has been received. When this condition is satisfied, the \emph{Rung condition-in} is set to \emph{True} as shown in Figure~\ref{msg_queue_fig}.  

\underline{\emph{Condition 2:}} The message queue has an empty slot. That is, the \emph{MSG.EW} bit is set to \emph{ON} as shown in Figure~\ref{msg_queue_fig}.

\end{minipage}}
\vspace{0.2cm}


\emph{Condition~1} means that the response for the last message has been received by the PLC. This process can take multiple \emph{scan cycle} times. An example analysis to find the time it takes to get a response for a previously sent message, is shown in Table~\ref{responseTime_table}. 
This data is based on experimental setup using SWaT testbed. The response time is calculated after a request has been received by the destination PLC and then the corresponding response has arrived at the source PLC. 
In Table~\ref{responseTime_table}, it can be seen that the response time from PLC~2 to PLC~3 is $11.062 ms$ on an average. 
From the same table, it can be seen that the \emph{scan cycle} time for PLC~3 measured using a system call is $4.117 ms$. This means that it takes multiple scan cycles to get the response back to PLC~3. When the response has arrived the \emph{Rung condition-in} in Figure~\ref{msg_queue_fig} becomes \emph{True}. The message instruction is scanned each \emph{scan cycle} but it does not get executed if \emph{Rung condition-in} is not set to True. In Figure~\ref{msg_queue_fig}, the black circle with a number at the bottom represents a \emph{scan cycle} count. At the first \emph{scan cycle}, when \emph{Rung condition-in} becomes True, MSG.EN~(message enable) bit is set to ON. Then the message is ready to enter the message queue and it checks MSG.EW~(message enable wait) bit and if it is full the message keeps waiting until MSG.EW is set to ON and message can enter the queue and get transmitted on the network~\cite{allenbradley_MSG_instruction_queuing}. As shown in Figure~\ref{msg_queue_fig}, it can take several \emph{scan cycle} times to complete the whole process. Therefore, the two mentioned conditions must be fulfilled to transmit the messages between PLCs. Since everything is measured in terms of \emph{scan cycle}s,  it would be possible to recover this \emph{scan cycle} information from the network layer and use it as a hardware and software fingerprint for each PLC.

\begin{table*}
\begin{center}
 \begin{tabular}[!htb]{|l | l| c | c | c | c | c |} 
 \hline
 PLC~X & MSG & E[$T_{SC}$] & E[$T_{Resp}$] & E[$T_{ESC}$] & E[$\eta$] = $\frac{E[T_{ESC}]}{E[T_{SC}]}$ & E[$\eta$] = $\frac{E[T_{ESC}]}{E[T_{SC}]}$  \\ 
 To PLC~Y & Instruction & (ms) & (ms) & (ms) & if $T_{SC} << T_{OH}$ & if $T_{SC} >> T_{OH}$  \\ 
 \hline
 PLC~1:  & FIT-201 & 4.387 & 8.597 & 22.153 & 5.34 & 1.96\\
 PLC~2 & IP: x.x.1.20 & & & & &  \\
 \hline

 PLC~2:  & LIT-301 & 4.708 & 8.273 & 34.327 & 7.04 & 1.98\\
 PLC~3 & IP: x.x.1.30 &    & & & &   \\
 \hline
 
  PLC~3:  & MV-201 & 4.117 & 11.062 & 39.814 & 9.75 & 1.97 \\
 PLC~2 & IP: x.x.1.20 &    &   & & & \\
 \hline

  PLC~4:  & MV-501 & 4.045 & 18.905 & 44.529 & 10.04 & 1.96\\
 PLC~5 & IP: x.x.1.50 &  & & & & \\
 \hline
 
  PLC~5:  &  FIT-401 & 5.078 & 12.804 & 57.126 & 11.08 & 1.97\\
 PLC~4 & IP: x.x.1.40 &  & & & & \\
 \hline
 
   PLC~6:  & LIT-101 & 2.721 & 3.13 & 25.364 & 9.32 & 1.97\\
 PLC~1 & IP: x.x.1.10 &  & & & & \\ [1ex]
 \hline

\end{tabular}

\vspace{1em}
\caption{MSG Instructions From/To PLCs and their respective timing analysis. $E[\cdot]$ is the expected value of a particular variable. $T_{SC}: Scan Cycle$, $T_{Resp}: MSG Response Time$, $T_{ESC}: Estimated Scan Cycle$, $MV: motorized valve, FIT: flow meter, LIT: level sensor$.  $\eta$ is the ratio between a \emph{scan cycle} and estimated \emph{scan cycle} time, bounds for that are derived in Apeendix~\ref{lower_bound_eta_appendix}}\label{responseTime_table}
 \vspace{-2.5em}
\end{center}
\end{table*}

\subsection{Threat Model}

An attacker can compromise a plant either by remotely entering the control network, physically damaging the PLCs/components of PLCs, or intercepting traffic as man-in-the-middle~(MiTM). It is assumed that the adversary aims to  sabotage a plant by compromising the communication between the PLCs and from PLCs to other devices such as  HMI, SCADA or historian server. Once an intrusion has happened an adversary can choose to spoof the messages by using fake IDs, suspend the messages~(denial of service) making the PLCs unavailable, intercepting and modifying the traffic, e.g., MiTM attack or a masquerade attack by suspending a legitimate PLC and sending fake messages on behalf of the legitimate PLC, which will ultimately falsify the current plant's state and lead potentially to unsafe states. The following attack scenarios are considered: Denial of Service~(DoS), Man-in-the-Middle~(MiTM), and Masquerading. Note that a masquerading attack can be realized  by an MiTM attack that also drops the original packets produced by a given PLC. 

\noindent For the masquerading attack three types of attackers are considered. \emph{Naive}, which tries to imitate a PLC but has no knowledge about the estimated \emph{scan cycle} of the PLC, \emph{Powerful Partial Distribution Knowledge~(PDK)}, which tries to imitate a PLC and knows the mean of the estimated scan cycle of a PLC, and the third case of \emph{Powerful Full Distribution Knowledge~(FDK)}, which tries to imitate a PLC and knows the full distribution of the estimated \emph{scan cycle}. A customized python script was developed using pycomm library to get precise control of message transmission during the masquerading attack.

\begin{figure*}[t]
\centering
\includegraphics[height=7cm,width=14cm]{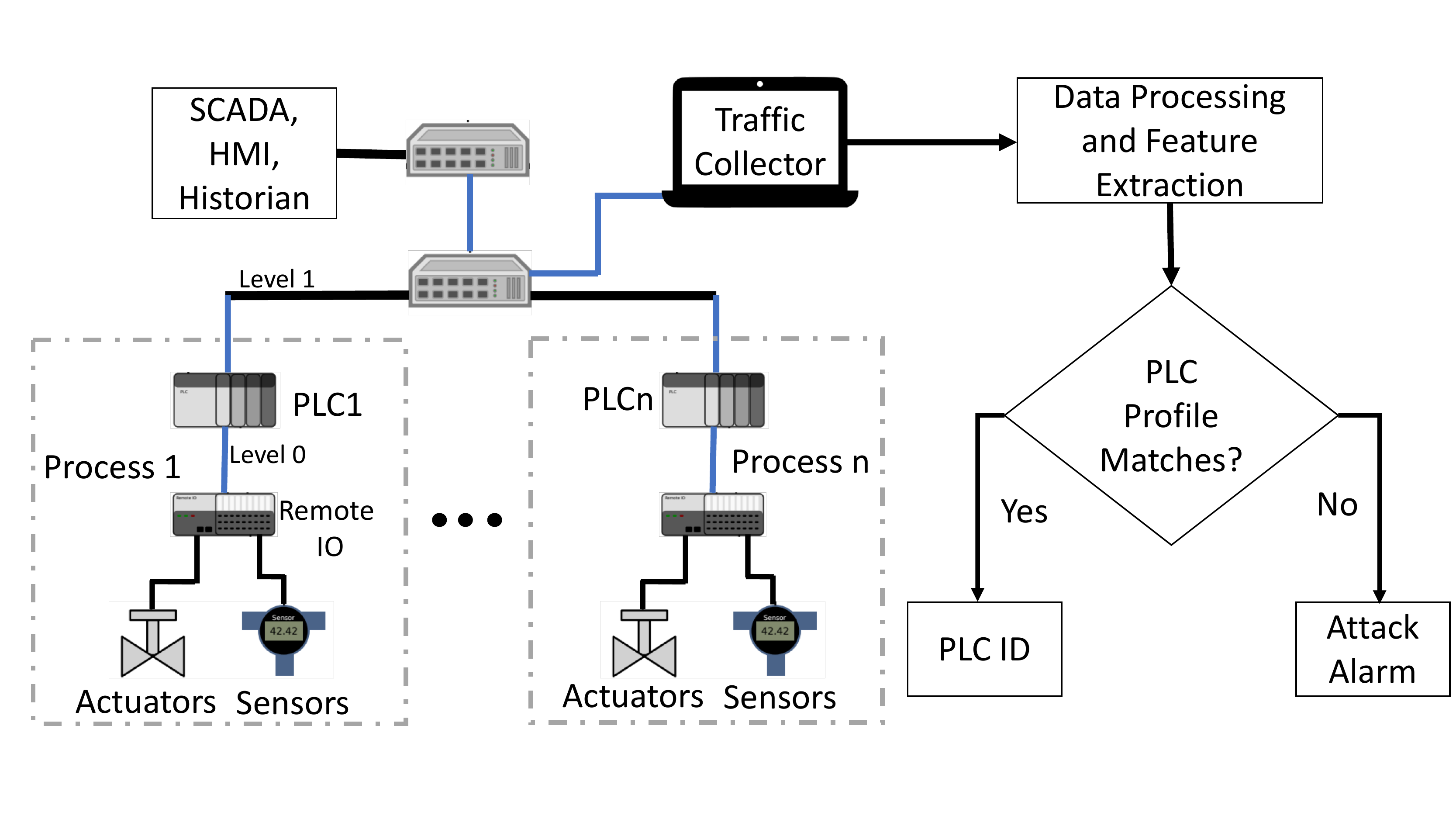}
\caption{Overview of the proposed technique.}
\label{framework_overview_fig}
\vspace{-1.5em}
\end{figure*}

\section{Design considerations}
\label{overview_sec}


\noindent \textbf{Problem Statement:}
The  following question is addressed in this work: {\em Is it possible to authenticate messages from each PLC in a non-intrusive and passive manner?}

\noindent \textbf{Proposed solution:} The idea is to create fingerprints for the PLCs based on hardware and software characteristics of the PLCs.  Once the PLCs have been identified the next step is to detect a range of network attacks on the Level~1 network communication. Figure\,\ref{framework_overview_fig} shows an overview of the proposed framework for device identification and attack detection. The proposed technique begins with network traffic data collection. The collected data is processed to estimate the \emph{scan cycle} time. Next, the estimated \emph{scan cycle} time is used to extract a set of time and frequency domain features. Extracted features are combined and labeled with a PLC ID. A machine learning algorithm is then used for PLC classification and attack detection.
 The traffic collector is deployed at the Level~1 network~(also known as SCADA control network~\cite{urbina_CCS2016limiting}) switch with mirror port to monitor all the network traffic at Level~1. Data is collected for all the six PLCs deployed in the SWaT testbed. 
The list of requested messages together with the requesting and responding PLC is given in Table~\ref{responseTime_table}.  

PLCs are profiled using the time and frequency domain features of the estimated \emph{scan cycle} samples.  Fast Fourier Transform\,\cite{welch1967} is used to convert data to the frequency domain and extract the spectral features. A list of all the features along with the description is presented in Table\,\ref{features} in Appendix~\ref{metrics_appendix} due to limited space. 
Data re-sampling is done to find out the sample size with which high classification accuracy can be achieved. This, in turn, would inform us about the time the proposed technique needs to make a classification decision.

\textit{Experimental Evaluation}:
\label{evaluation_sec}
The experiments are carried out in a state-of-the-art water treatment facility~\cite{swat2016} and a smart grid testbed. The proposed technique is tested on six different Allen Bradley PLCs, four Siemen IEDs and four Wago PLCs. 

\begin{figure*}[htb]
\centering
\includegraphics[height=6 cm,width=15cm]{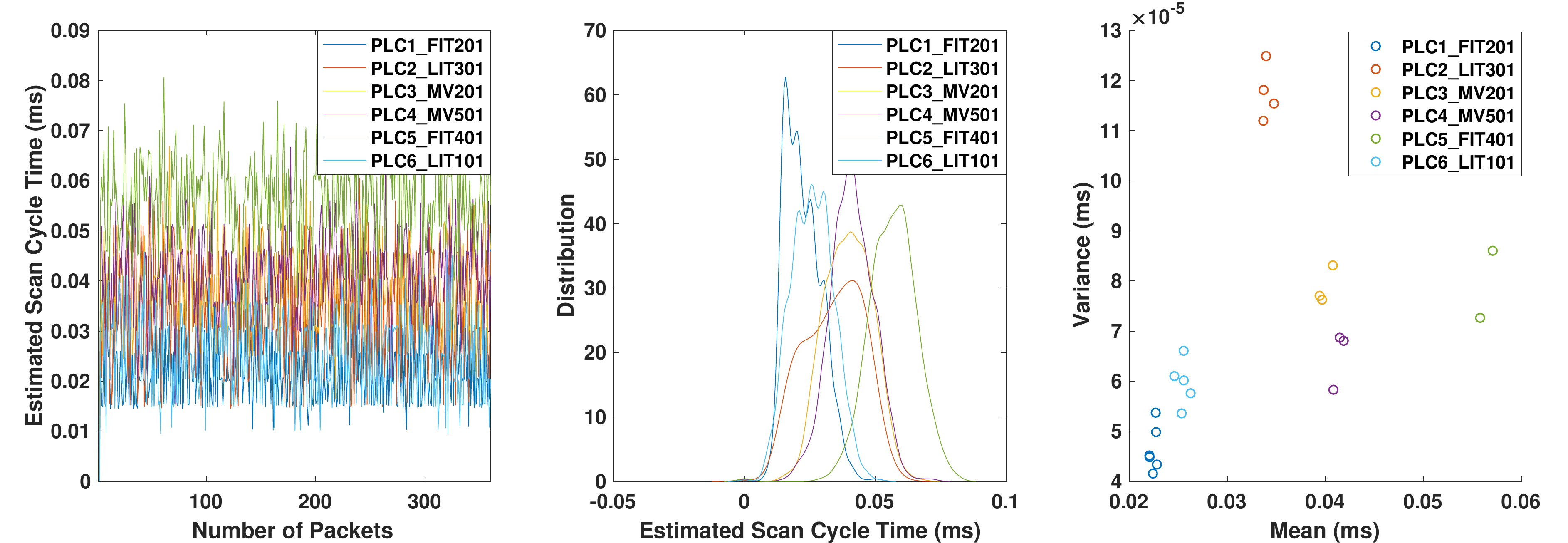}
\caption{ Leftmost plot shows estimated \emph{scan cycle} times for six PLCs. In the middle, the distribution for estimated \emph{scan cycle} is shown. On the right, two time-domain features are used to classify the PLCs from each other.}
\label{proof_of_fingerprint_fig}
\vspace{-1.5em}
\end{figure*}

\subsection{Identification of PLCs}

As shown in Figure~\ref{framework_overview_fig}, the proposed technique compares a PLC data with a pre-created model and if the profile is matched it returns the PLC ID. 
During the testing phase if the profile does not match to the pre-trained model an alarm is raised and a potential attack is declared. Experimental results are presented in the following in the form of the research questions. 

\emph{RQ1: Proof of fingerprint. Is the estimated \emph{scan cycle} a good candidate for a fingerprint?} Figure~\ref{proof_of_fingerprint_fig} shows statistical features of the estimated \emph{scan cycle} vector for six PLCs in the SWaT testbed. On the leftmost plot, the time series data of the estimated \emph{scan cycle} time is plotted. It can be observed from the middle plot that the distributions of the estimated \emph{scan cycle} time of the PLCs have distinctive behavior but still a few PLCs overlap. The rightmost plot shows two time-domain features namely mean and variance of the estimated \emph{scan cycle} time. By using these two features, the six PLCs can be easily distinguished. This visual representation is a proof for the existence of \emph{scan cycle} based fingerprint. These two features are used for the ease of visualization, however, to see the performance of the proposed technique the dataset is analyzed in a systematic manner using machine learning and the complete feature set. The results are shown in Table~\ref{chunksize_vs_accuracy_table} and discussed in the following.

\emph{RQ2: PLC Identification/Attack Detection Delay. What is the right amount of data to identify PLCs with the highest accuracy?}  It is observed that $120$ samples are a good trade-off between accuracy and detection time with an accuracy of $96.12\%$. On average it takes just $3.6$ seconds to make a detection decision.  


\emph{RQ3: How will the amount of training and testing data affect PLC identification performance?}  Results in Table~\ref{kfold_cross_validation_table}, point out that the accuracy of our chosen classifier function is stable for the range of data divisions and does not depend on the choice of size of the dataset. This gives a practical insight into the case when limited data is available to train the machine learning model.

\emph{RQ4:  Is the fingerprint stable for different runs of the experiments?} The data collected at $22$ degree Celsius in scenario~1 is used to train a machine learning model and test it with the data collected in scenario~2 at $33$ degree Celsius. The first row in Table~\ref{two_class_table} shows the results for this experiment. 
Table~\ref{two_class_table} ensures that the fingerprint is stable for different runs and temperature variations. The use of the binary classifier also helps in arguing about the scalability of the proposed technique, a topic of the following research question.



\begin{table}
\begin{center}
\caption{Multiclass Classification: chunk size vs classification accuracy.}
\label{chunksize_vs_accuracy_table}
\begin{adjustbox}{max width=0.5\textwidth}
 \begin{tabular}[!htb]{|c | c| c | c | c| c| c | c |} 
 \hline
 Chunk Size & 10 &30  &  60  &  100  & \textbf{120}  &  150  &  200  \\ 
 \hline
  Accuracy &  59.8854\% & 82.4561\% &  87.2727\%  & 93.5484\% & \textbf{96.1538}\% &90.4762\% & 92.8571\% \\ [1ex]
 \hline
 
\end{tabular}
\end{adjustbox}
\end{center}
\end{table}


\begin{table}
\begin{center}
\caption{k-fold cross validation using multi-class classifier.}
\label{kfold_cross_validation_table}
\begin{adjustbox}{max width=0.5\textwidth}
 \begin{tabular}[!htb]{|c | c| c | c | c| c| c |} 
 \hline
 k-fold  & 2 & 3 & 5 & 10 & 15 & 20   \\ 
 \hline
 Accuracy  &  88.2883\% &  85.5856\% &  89.1892\% &  88.2883\% &  89.6396\% &  90.0901\% \\ [1ex]
 \hline
 
\end{tabular}
\end{adjustbox}
\end{center}
\end{table}

\begin{table}
\begin{center}
\caption{Data from each PLC is labeled as class-1 and all the data from rest of the five PLCs are labeled as class-2. Results show stability from run-to-run and across temperature range.}
\label{two_class_table}
\begin{minipage}{10cm}

 \begin{tabular}[!htb]{| c | c| c | c | c| c| c | } 
 \hline
  & PLC~1 &PLC~2  &  PLC~3  &  PLC~4  & PLC~5  &  PLC~6   \\ 
 \hline
 
    Stable \footnote{\centering Stability under temperature variation, scalability using  2-class classifier  } &  99.03\% & 100\% &  84.47\%  & 87.38\% & 100\% & 92.23\%  \\
     \hline

  Scale \footnote{\centering Scalability using the one class classifier approach} &  94.44\% & 97.22\% & 93.33\%  & 96.43\% & 94.44\% & 93.33\%  \\ [1ex]
 \hline
 
\end{tabular}
\end{minipage}
\end{center}
\end{table}

\emph{RQ5: With an increase in the number of PLCs, how accurately PLCs can be classified?} At this point a different line of argument is considered, that is, it is not necessary to compare hundreds of PLCs with each other. Since the source~(expected) PLC of a message is known, the job is to verify if this message is really being generated by that particular PLC or being sent by an attacker device or being spoofed at the network layer. In Table~\ref{two_class_table} first row shows the accuracy for PLC identification based on this binary classification. Another important question is, can attacks be detected using such supervised machine learning models? To resolve this problem, SVM is used as a one-class classifier. In Table~\ref{two_class_table} the second row shows the results with one-class SVM~(OC-SVM), to identify a particular PLC resulting in higher accuracy. Using OC-SVM makes the argument of scalability even stronger since a model can be created by using just the normal data of a PLC. In the following section, the performance of attack detection using one-class classification model is discussed.

\begin{table}
\begin{center}
\caption{Accuracy improvement due to change in the control logic of PLC~4. DF: Default Profiles, MF: Modified Profiles.}
\label{accuracy_improvement_table}
\begin{adjustbox}{max width=0.48\textwidth}
 \begin{tabular}[!htb]{|c | c| c | c | c| c| c | c | } 
 \hline
  & PLC~1 &PLC~2  &  PLC~3  &  \textbf{PLC~4}  & PLC~5  &  PLC~6  & \textbf{Accuracy}  \\ 
 \hline
 
  DF~(ms) &  22.31 & 34.07 &  39.71  & \textbf{41.08} & 56.78 & 25.82 & \textbf{93.54}  \\
\hline
  MF~(ms) &  22.15 & 34.32 & 39.81  & \textbf{44.58} & 57.12 & 25.36 & \textbf{99.06\%}  \\ [1ex]
 \hline
 
\end{tabular}
\end{adjustbox}
\end{center}
\end{table}


\begin{table}
\begin{center}

 \begin{tabular}[!htb]{|c | c| } 
 \hline
 Device Type & Identification Accuracy    \\ 
 \hline
  4 x Siemen IEDs &   82.2148\%  \\ 
 \hline
 4 x Wago PLCs & 70.0909\%       \\  [1ex]
 \hline
 
\end{tabular}
\end{center}
\caption{EPIC Testbed Performance Evaluation. Table shows the results for a multi-class classification.} 
\label{EPIC_testbed_table}

\end{table}

\subsection{Practical application of the proposed technique in an ICS} \label{practical_app_subsection}
One unique feature of the proposed technique is that it is the combination of PLC hardware and control logic execution time. Therefore, it is possible to create a unique fingerprint even for similar PLCs with the similar control logic which is probable in an industrial control system. In the Table~\ref{chunksize_vs_accuracy_table}, it was seen that the multiclass accuracy to uniquely identify all the six PLCs in SWaT testbed is $93.54\%$ for a sample size of $100$. From the Figure~\ref{proof_of_fingerprint_fig} it was observed that PLC~3 and PLC~4 have a very similar profile hence lower classification accuracy. Considering that the fingerprint is the combination of the hardware and the control logic, an experiment to force a distinguishing fingerprint is proposed. To remove the collision between the PLC~3 and PLC~4, an extra delay was added to the ladder logic of the PLC~4 without affecting the normal operation. The classification accuracy for multiclass classification of six PLCs is increased to $99.06\%$. The results are summarized in Table~\ref{accuracy_improvement_table}.

\subsection{Generalising to Other ICSs and Devices}

The proposed technique has been tested in another physical process (electric power grid testbed) EPIC\cite{Nandha_ESORICS2020_EpicDataset} employing different type of devices (WAGO PLCs and Siemens IEDs (Intelligent Electronic Devices)). The EPIC is divided into four main sectors: Generation, Transmission, Micro-Grid and Smart Home. Each sector comprised of various electrical equipment such as motors, generators and load banks. These equipment can be monitored and managed by different digital control components such as Programmable Logic Controllers (PLC), Intelligent Electronic Device (IED) through different communication medium. Generic Object Oriented Substation Event (GOOSE), MODBUS Serial, TCP/IP and Manufacturing Messaging Specification (MMS) are employed from the IEC 61850 standard communication networks and systems in substations~\cite{iec61850_overview}. Siemens Protection Relay Intelligent Electronic Device (IED) which provides Protection, Instrumentation and Metering functionality. We have used four Siemens 7SR242 series IEDs. This way we have diversified both the devices being used as well as the communication protocols.

Results are shown in Table~\ref{EPIC_testbed_table}. Timing profiles obtained for all four devices compared against the signature of each device and the accurate identification percentage is reported in Table~\ref{EPIC_testbed_table}. For the same four PLCs with identical control logic a 70\% device identification is result is encouraging. Moreover, it is to be noted from Section~\ref{practical_app_subsection} that by modifying the control logic in a benign manner it is possible to achieve accuracy close to 100\%.


\begin{figure}[t]
\centering
\includegraphics[height=5cm,width=9cm]{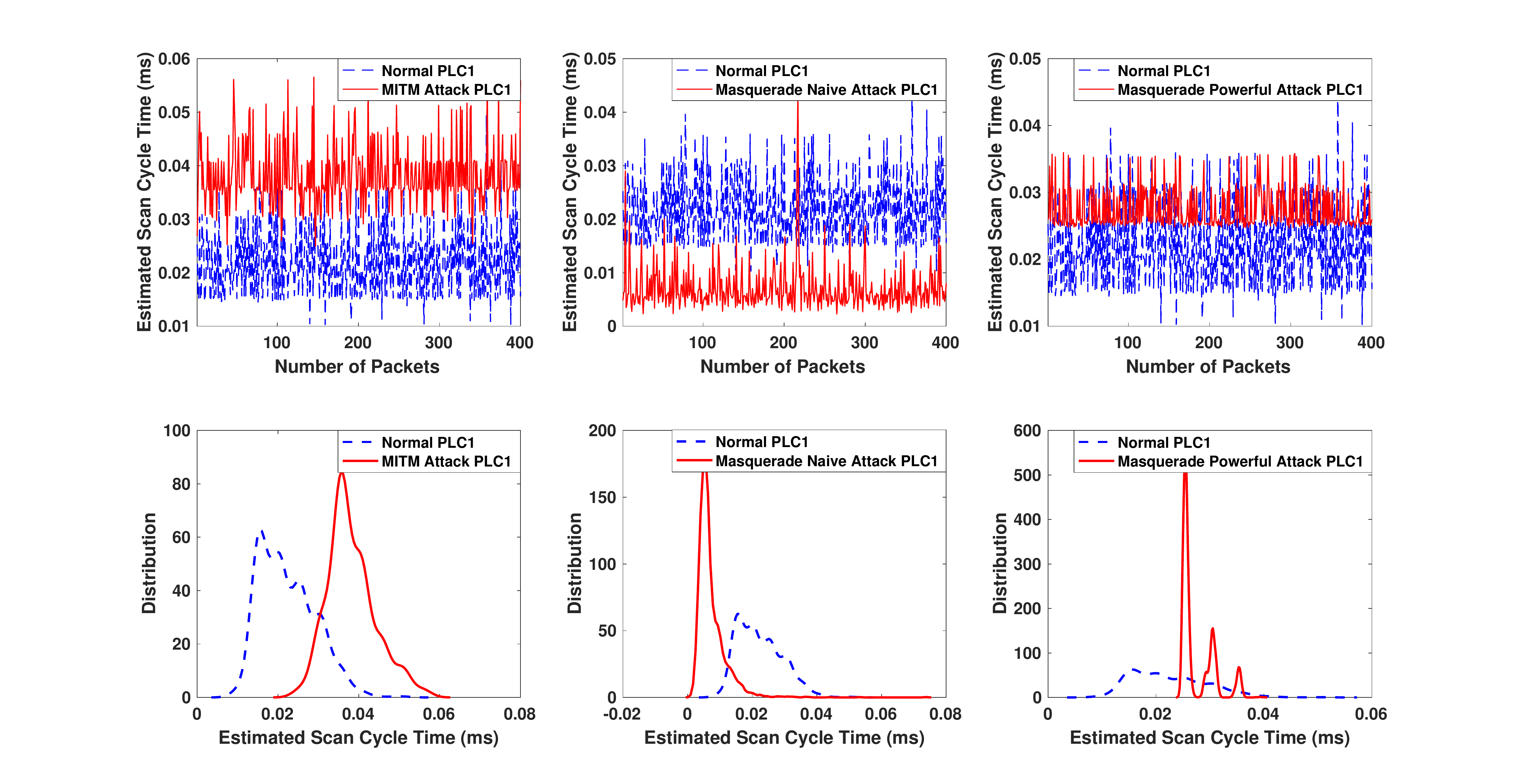}
\caption{MiTM and Masquerade Attack~(Naive and Partial Distribution Knowledge) scenario for PLC1.}
\label{plc1_attacks_fig}
\end{figure}


\begin{figure*}[t]
\centering
\includegraphics[height=9cm,width=20cm]{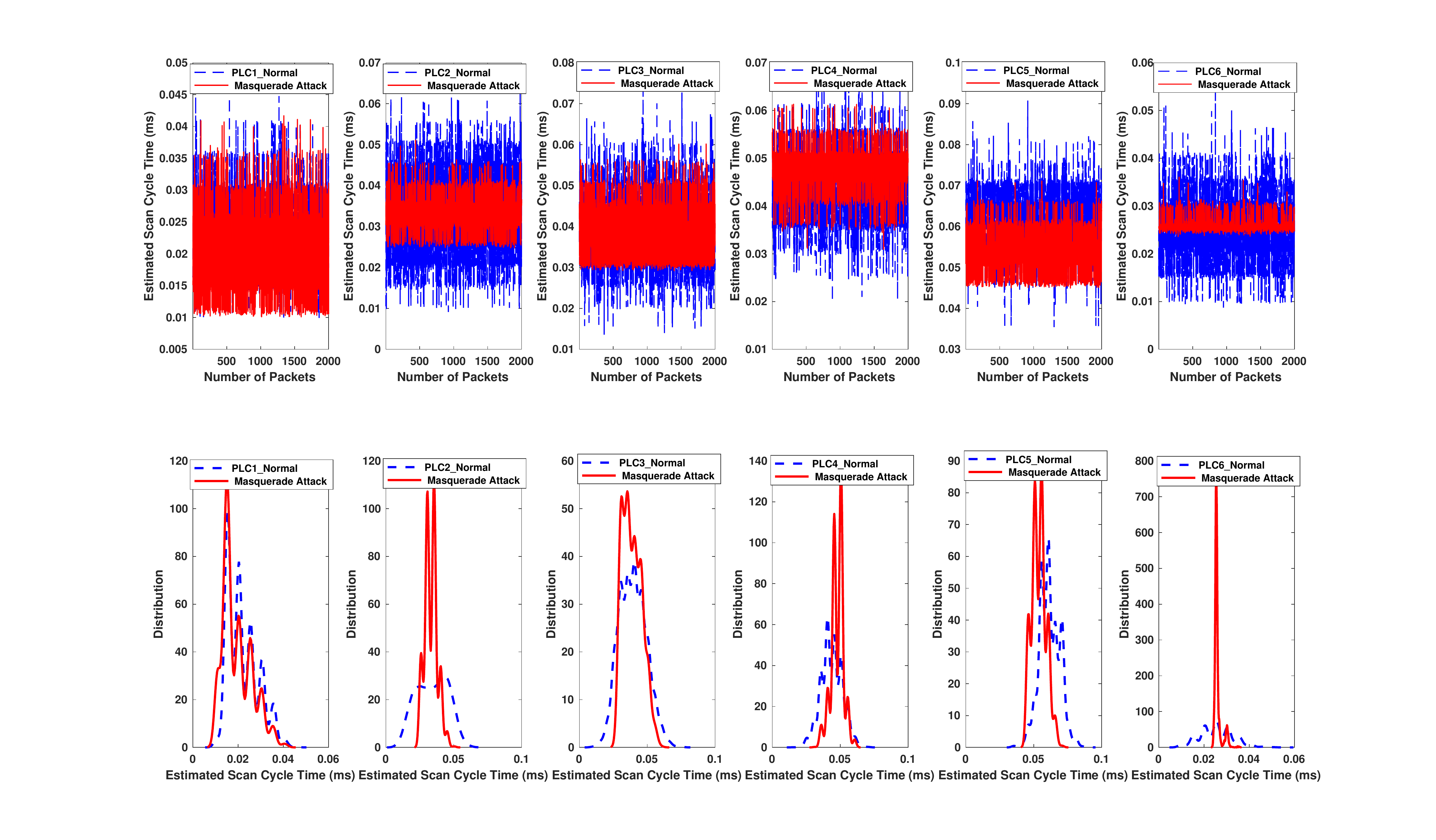}
\caption{Powerful Masquerade Attack~(Full Distribution Knowledge) on all six PLCs.}
\label{all_6plcs_masquerade_attack_fullDist_fig}
\end{figure*}


\begin{table}
\begin{center}

 \begin{tabular}[!htb]{|c | c| c | c | c| c |  } 
 \hline
 Attack Type & PLC~1 &PLC~2  &  PLC~3  & PLC~5  &  PLC~6   \\ 
 \hline
 
  MiTM: TPR &  100\% & 100\%   & 100\% & 100\% & 100\%  \\
 MiTM: FNR &  0\% & 0\%   & 0\% & 0\% & 0\%  \\
     \hline
   MN TPR: &  100\% & 100\% &  100\%  & 100\%  & 100\%  \\
   MN FNR &  0\% & 0\% &  0\%  & 0\%  & 0\%  \\
   \hline

  MPDK TPR: &  100\% & 100\%   & 22\% & 100\% & 100\%  \\ 
  MPDK FNR: &  0\% & 0\%   & 78\% & 0\% & 0\%  \\
   
      \hline

  MFDK TPR: &  0\% & 82.35\%   & 0\% & 
  100\%  & 100\%\\ 
  MFDK FNR: &  100\% & 17.64\%   & 100\% 
   & 0\%  & 0\%\\ [1ex]
 \hline
 
\end{tabular}
\vspace{1em}
\caption{Attack Detection Performance. MN: Masquerader Naive, MPDK: Masquerader Partial Distribution Knowledge, MFDK: Masquerader Full Distribution Knowledge.}
\label{mitm_imposter_attack_detection_table}
\vspace{-2em}
\end{center}
\end{table}

\subsection{Attack Detection}

A powerful masquerader with the knowledge of the network traffic pattern can try to maintain the normal network traffic statistics. Such a masquerade attacker would deceive the network traffic based intrusion detection methods~\cite{Mitchell_2014_IDS_CPS_survey}. The proposed technique is based on the hardware and software characteristics of the devices, which are hard for an attacker to replicate~\cite{dey-2014}. 

\emph{RQ6: How well does the proposed technique perform to detect powerful network attacks on PLCs?}  

The intuition behind the attack detection is that an attack on the network even if not affecting the network traffic statistics, must cause  deviation to the estimated \emph{scan cycle} fingerprint profile of the associated PLC. In Table~\ref{mitm_imposter_attack_detection_table} the attack detection rate~(TPR) and attack missing rate~(FNR) are shown. The first row shows the attack detection performance for a MiTM attacker which intercepts the traffic between the PLCs and then forwards the compromised messages. It is observed that all the attacks on all PLCs are detected with $100\%$ accuracy. To understand these attack scenarios  consider Figure~\ref{plc1_attacks_fig} that contains plots for the estimated \emph{scan cycle} for PLC~1 under attack and normal operation. 
For the masquerading attack three types of attackers are considered. \emph{Naive}, which tries to imitate a PLC but has no knowledge about the estimated \emph{scan cycle} of the PLC, \emph{Powerful Partial Distribution Knowledge~(PDK)}, which tries to imitate a PLC and knows the mean of the estimated scan cycle of a PLC, and the third case of \emph{Powerful Full Distribution Knowledge~(FDK)}, which tries to imitate a PLC and knows the full distribution of the estimated \emph{scan cycle}. 

A powerful masquerader tries to imitate a PLC by sending  fake messages at  the exact time using its knowledge. Now this powerful attacker could not be detected by the network traffic pattern based methods because the number of packets, packet length, header information and other network profiles would all be the same as normal operation. Our proposed technique is able to detect this attack because the attacker deviates from the fingerprinted profile. In the rightmost plot in Figure~\ref{plc1_attacks_fig} it can be seen that the profile under this masquerading attack deviates massively from the normal fingerprint profile although the number of packets and other network configurations are not that different. This result is reflected in Table~\ref{mitm_imposter_attack_detection_table} in the third row where except one case all the attacks are detected with $100\%$ accuracy. Amid the high accuracy, one can make the attacker even more powerful by providing it with the complete distribution of the estimated \emph{scan cycle} vector. Row~$4$ in  Table~\ref{mitm_imposter_attack_detection_table} contains results for such an attacker. For  PLC~1 and PLC~3 the attacker was able to imitate the PLCs  perfectly thus  avoiding detection. Even though attacker has the full distribution knowledge  but still due to attacker's hardware imperfections for some scenarios  high  detection rates were obtained. To reinforce this result see  Figure~\ref{all_6plcs_masquerade_attack_fullDist_fig} which shows all the PLCs under this powerful masquerade attack. Observe from the top row  how similar attacked data time series is to the normal data. This result is very significant in the sense that the attacker does not change the network statistics and sends the fake messages pretending to be one of the legitimate PLCs. It is the unique characteristic~(\emph{scan cycle}, queuing load) fingerprint of the PLC which allows attack detection.

\begin{figure}
    \centering
    \includegraphics[scale=0.35]{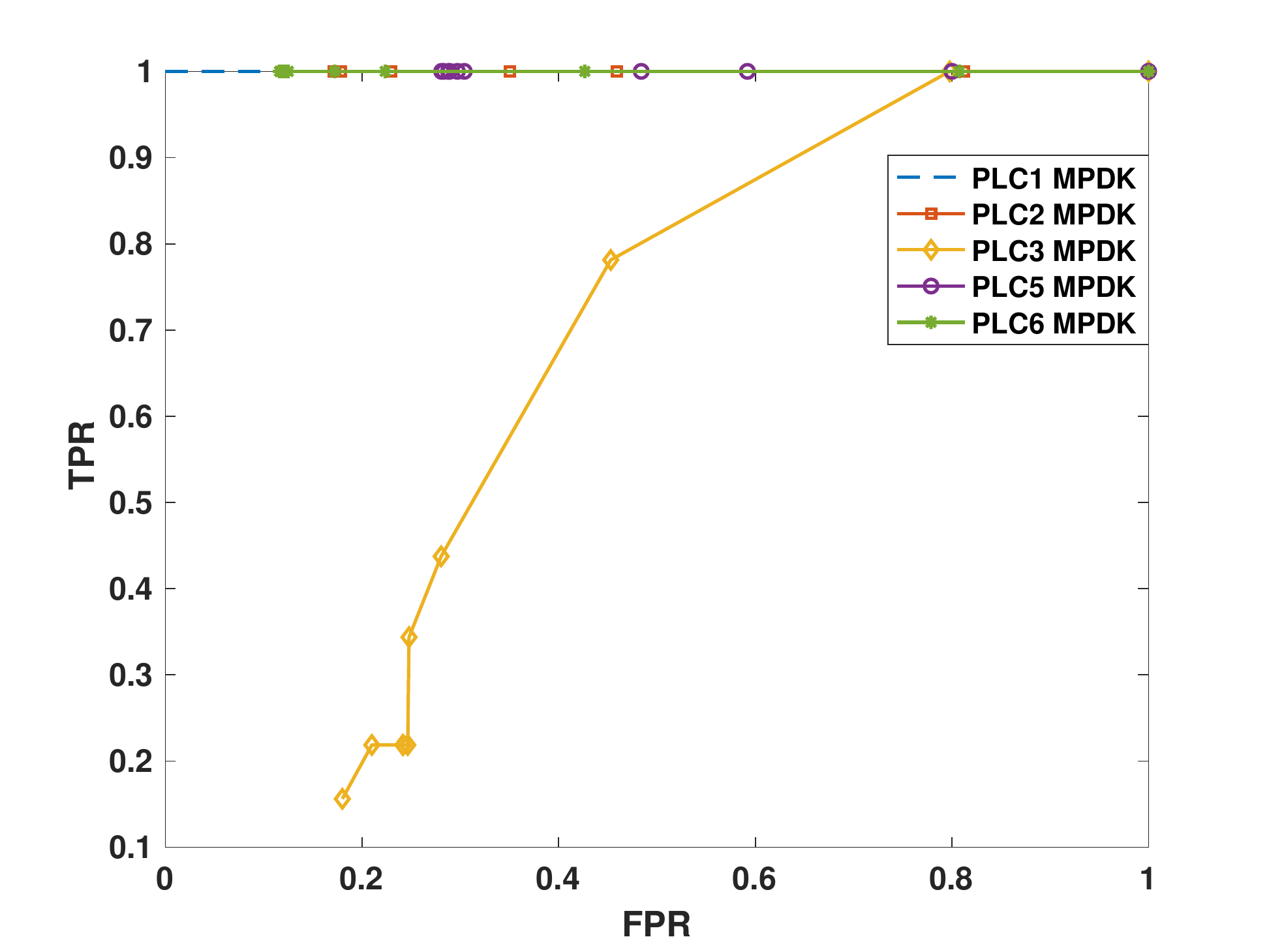}
    \caption{ROC for Masquerader Partial Distribution Knowledge.}
    \label{fig:mpdk_ROC}
    \vspace{-2em}
\end{figure}

\begin{figure}
    \centering
    \includegraphics[scale=0.35]{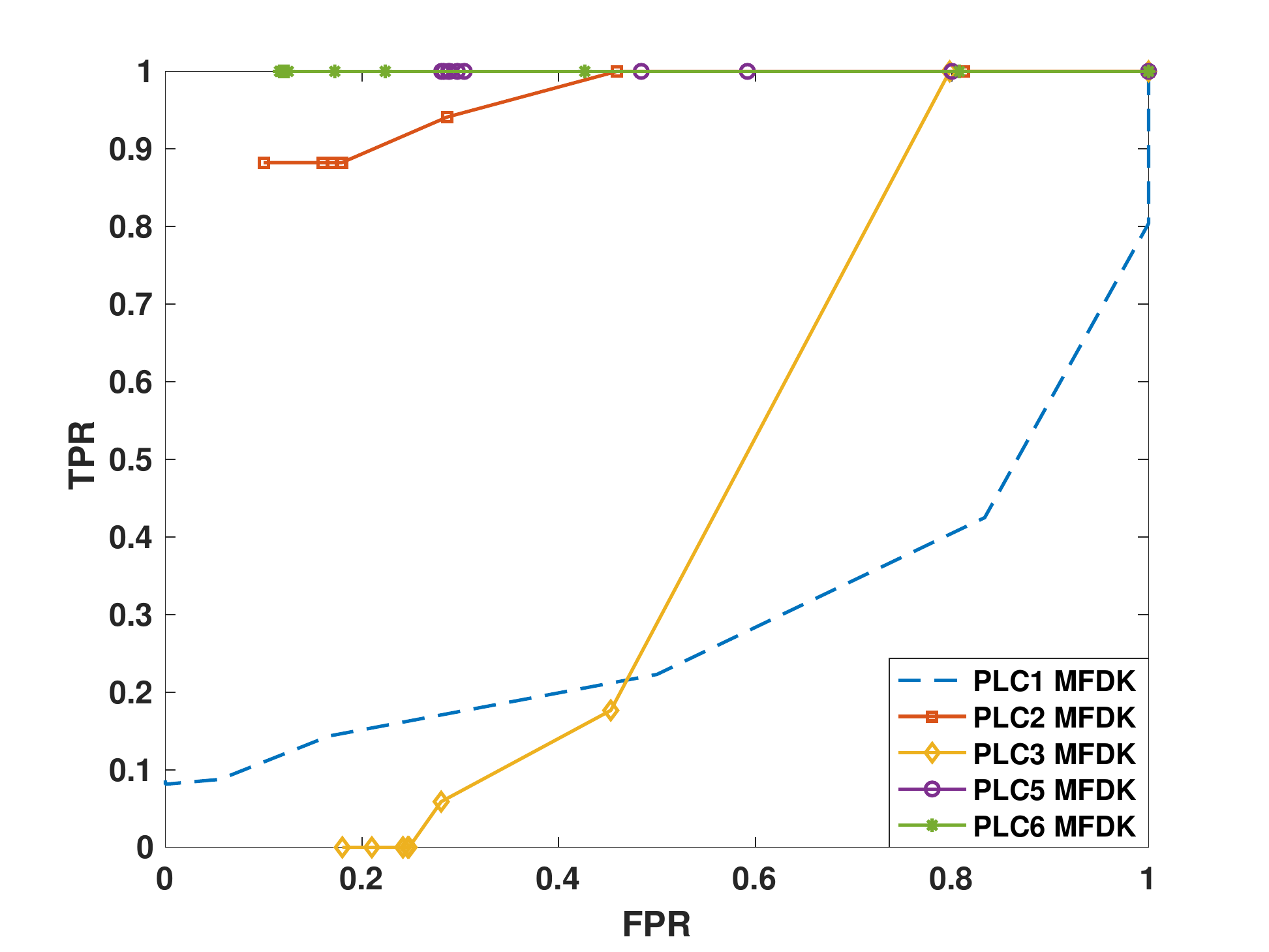}
    \caption{ROC for MFDK Masquerader Full Distribution Knowledge.}
    \label{fig:mfdk_ROC}
    \vspace{-2em}
\end{figure}

\emph{Performance evaluation of the classifier}:
A one-class SVM is used to detect attacks. To visualize the performance of the classifier a Receiver Operating Curve (ROC) is plotted using TPR (attacks rightfully detected) and FPR (normal data detected as an attack). 
SVM model measures the confidence that the test data belongs to the data on which model was trained~\cite{microsoft_oneclassSVM_platt1999estimating}. 
ROC plot for the last two attack scenarios from Table~\ref{mitm_imposter_attack_detection_table} are shown in Figure~\ref{fig:mpdk_ROC} and Figure~\ref{fig:mfdk_ROC} respectively. 
From Figure~\ref{fig:mfdk_ROC} it can be seen that for PLC~1 and PLC~3 if we are willing to accept a very high FPR then the attack detection could also be made. For PLC~2 the detection rate was not 100\% and hence as seen in Figure~\ref{fig:mfdk_ROC} an increase in FPR can result in high TPR. The key takeaway from this analysis of the classifier is that our classifier is not trained for very high FPR to demonstrate the 100\% attack detection rate. High detection rate is the result of underlying proposed detection technique based on the estimated \emph{scan cycle} time.


It is observed from
Figure~\ref{all_6plcs_masquerade_attack_fullDist_fig}  that the attacker was able to imitate the message transmission behavior of a PLC but still got exposed in most cases due to the limitation in the attacker's own hardware, as have been reported for PCs earlier~\cite{gtid_raheem_device_fingerprinting_tdsc2015}. Nevertheless, we do not assume any limitations on attacker's part and consider that an attacker is capable of perfectly imitating the PLCs and is also able to do a replay attack. In the following, we present a novel \plcwm technique to detect intrusions from such powerful attackers.


\begin{figure}[t]
\centering
\includegraphics[height=5cm,width=10cm,keepaspectratio]{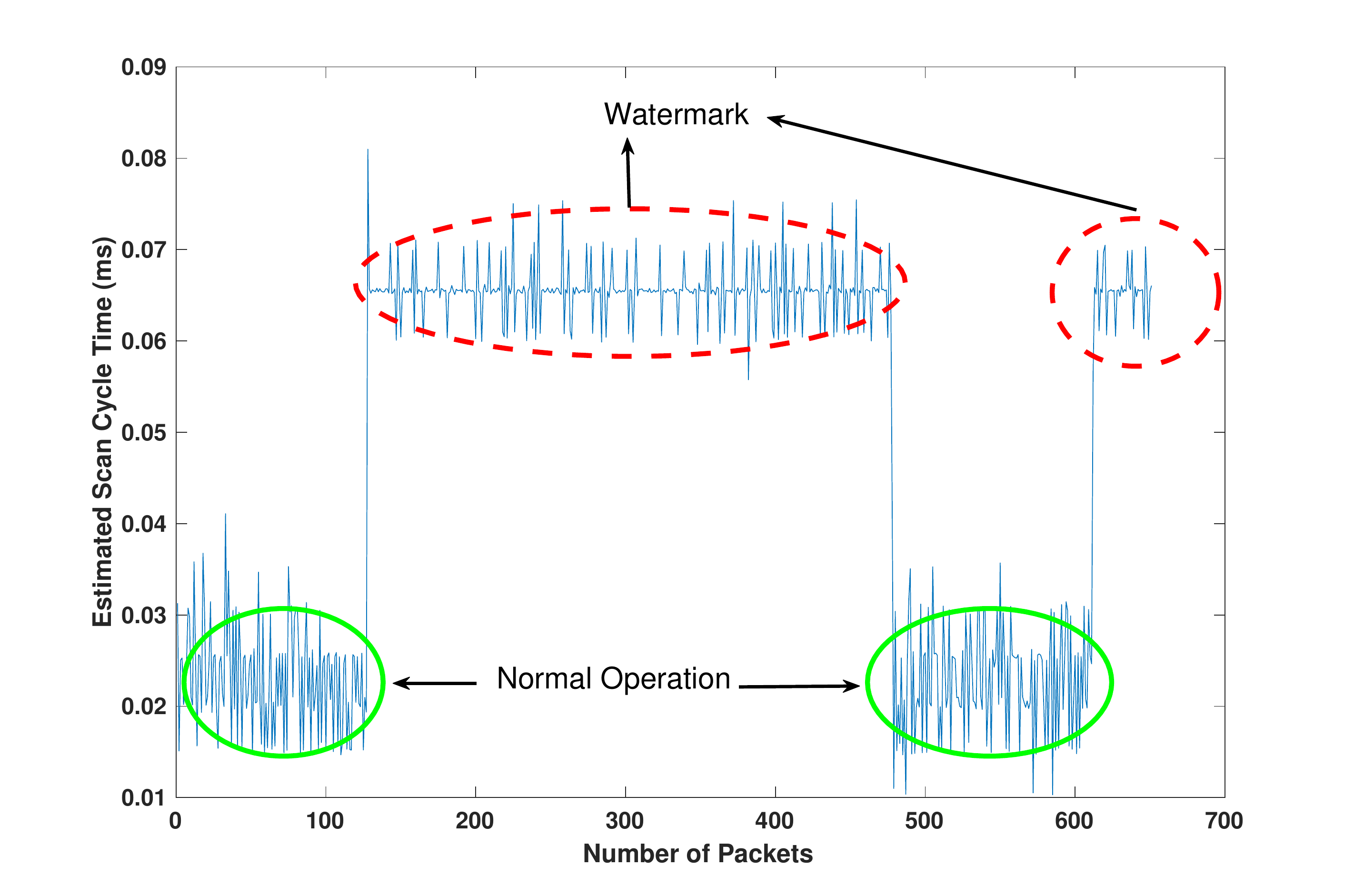}
\caption{Example watermark for  PLC~1 in SWaT.}
\label{CRP_PLC1_fig}
\vspace{-1.5em}
\end{figure}

\section{ PLC Waterkmarking}

 \plcwm exploits the relationship of PLC's unique feature of \sct  and the network layer data request messages exchanged among the PLCs. The idea of \plcwm is to extend a static fingerprint as discussed in the previous sections to a dynamic, randomly generated scheme to tackle a powerful attacker. Randomness in the watermark is generated by, 1)~using the clock of the PLC to inject random delay and 2)~injecting the watermark signal for a random number of scan cycle count, i.e., number of scan cycles for which a particular watermark is to be added.
 An example is shown in Figure~\ref{CRP_PLC1_fig} where a constant watermark signal is injected labeled as the watermark. The y-axis depicts the change in $T_{ESC}$ because of the watermark and x-axis  the change in the number of packets due to the watermark. 

\begin{figure}[t]
\centering
\includegraphics[height=4cm,width=8cm]{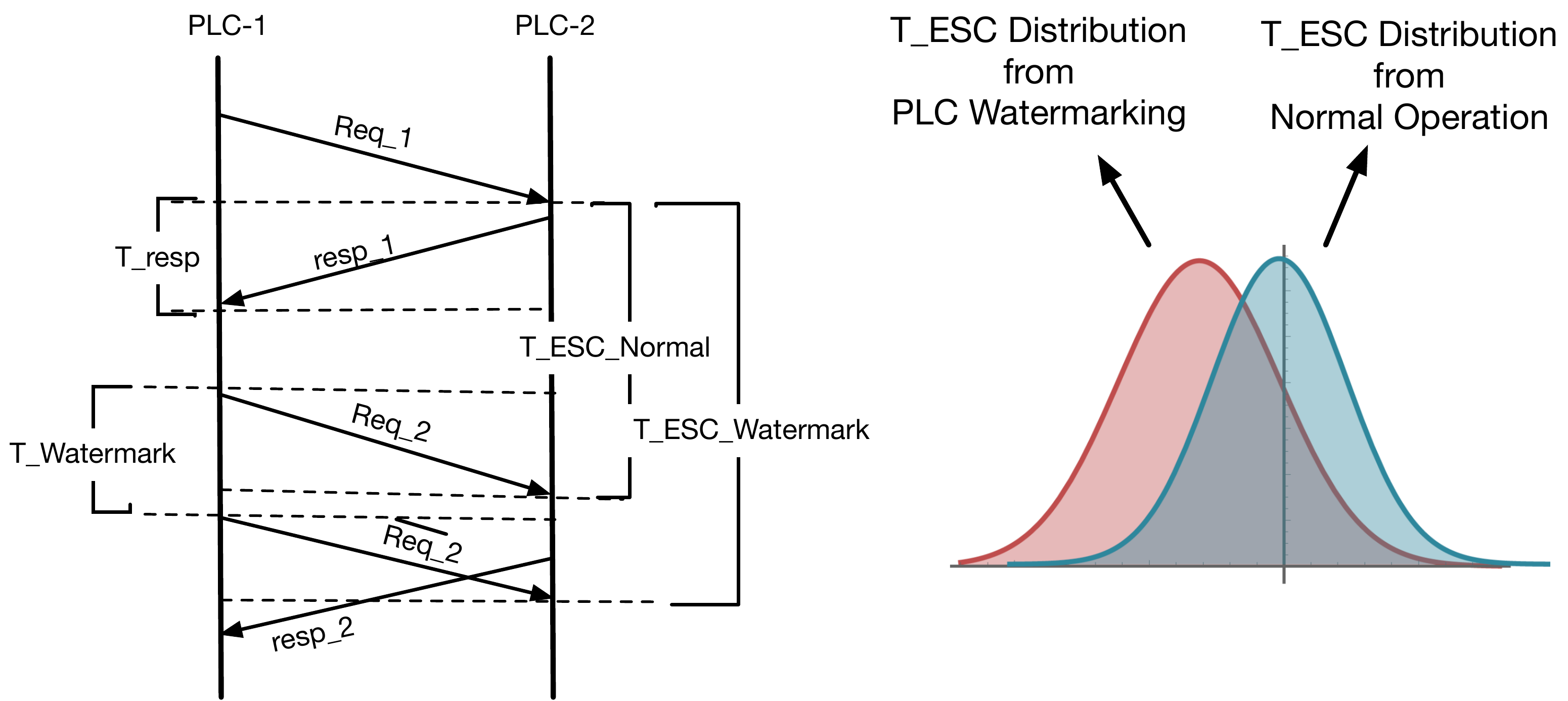}
\caption{Design of the \plcwm.}
\label{CRP_fig}
\vspace{-1em}
\end{figure}

\begin{figure}[t]
\centering
\includegraphics[height=9cm,width=7cm,keepaspectratio]{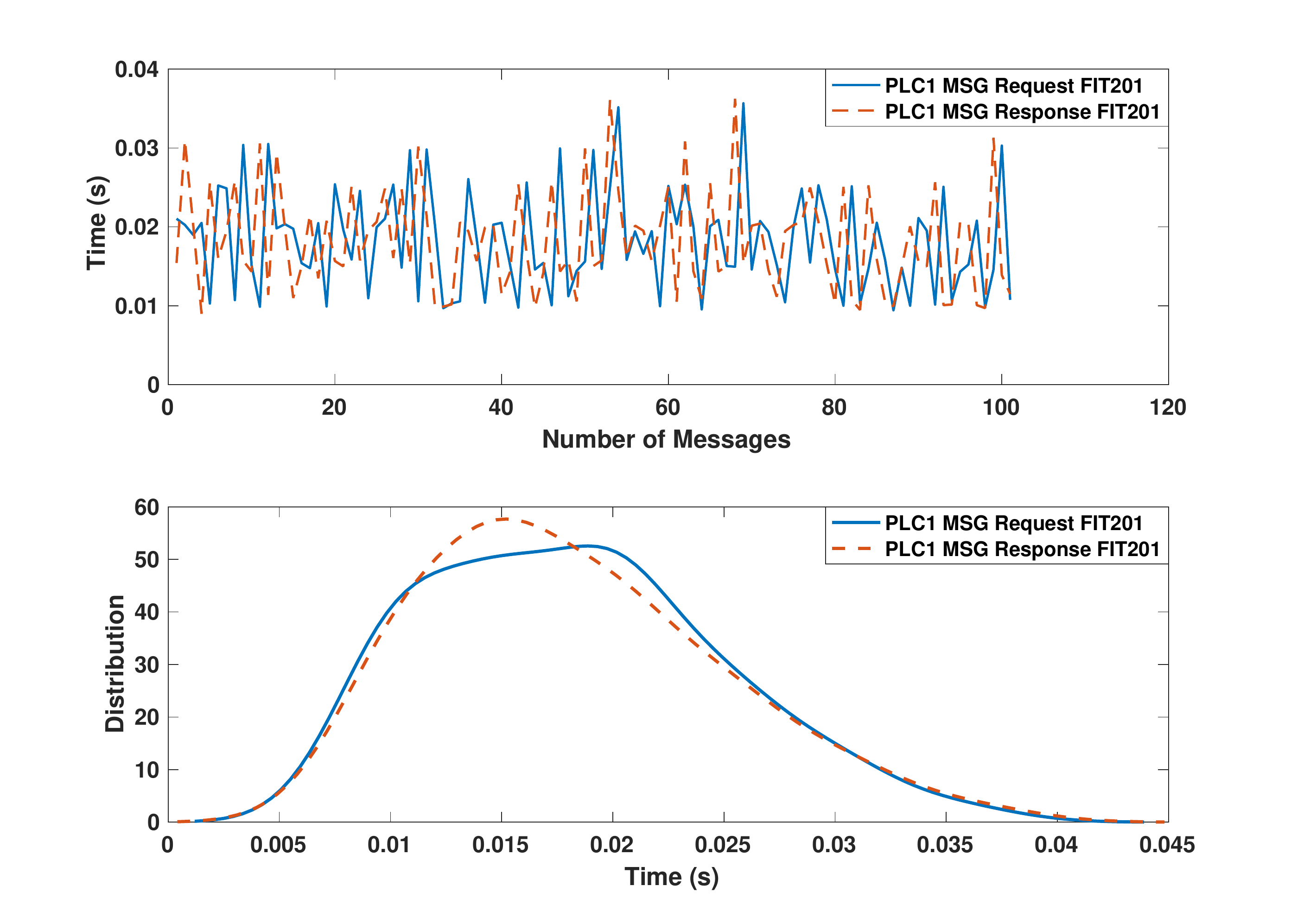}
\caption{Request-Response messaging in the PLCs.}
\vspace{-1.5em}
\label{req_resp_strong_Function_fig}
\end{figure}

The request-response model for data exchange among the PLCs facilitated the design of \plcwm. It has been observed that, 1)~the request messages are controlled by \sct and 2)~ the time of arrival of response messages is a function of request message arrival time. These two observations led us to, 1)~affect the message control by manipulating the \sct and 2)~exploit the feedback loop for request-response channel to inject and observe a watermark signal, respectively.  Figure~\ref{CRP_fig} presents the first hypothesis related to the \sct.  A watermark is injected at the end of the control logic (i.e., end of scan cycle) to introduce delays to the transmitted \emph{data request} messages on the network. Figure~\ref{req_resp_strong_Function_fig} shows the result for the second hypothesis depicting that the timing profile for the response messages has the similar distribution as the request messages. 

In Figure~\ref{CRP_fig}, an example of communication between two PLCs is shown. For simplicity, PLC~1 is assumed to transmit messages to PLC~2, representing explicit message exchange between PLCs. Under normal operation, PLC~1 sends a request to PLC~2 labeled as $Req_1$ and after some time gets the response labeled as $resp_1$. The second request is labeled as $Req_2$. The time between these two requests, and similar subsequent requests, establish a profile for estimated \emph{scan cycle} time as shown in previous sections. A random delay, labeled  $T_{Watermark}$, can be added making request~2 at a later time labeled as $\bar{Req_2}$. The time difference between $Req_1$ and $\bar{Req_2}$, and the subsequent packets using $T_{\text{Watermark}}$, constitutes a profile for  \plcwm.  The plot on the right-hand side in  Figure~\ref{CRP_fig} depicts the distributions for the case of the normal operation of the PLCs and with a watermark. An example of such a  watermark is shown in Figure~\ref{CRP_PLC1_fig} for the case of PLC1 in SWaT testbed. It is important to respect the real-time constraint also known as \emph{watchdog} timer: 
$[\max T_{SC}]_{Watermark} < T_{watchdog}$.


\subsection{Distinguishing Watermarked Signal from Normal}
In this section, the goal is to investigate  how the watermark could be distinguished from the normal profile, and how to build such a watermark. 
The result in Figure~\ref{req_resp_strong_Function_fig} shows a Gaussian approximation for \emph{estimated scan cycle time} $T_{ESC}$. Gaussian distribution possesses useful properties, e.g., scaling and shifting of a Gaussian random variable preserves its distribution.  


\begin{proposition}\label{gaussian_dist_proposition}
\emph{Linear transformation of a random variable}. For a random variable $\mathbf{X}$ with mean $\mu$ and standard deviation $\sigma$ as defined by a Gaussian distribution, then for any $a,b \in \mathbb{R}$ 
\begin{equation}
    \mathbf{Y} = a\mathbf{X} + b
\end{equation}
then, $\mathbf{Y}$ is a Gaussian random variable with mean $a\mu + b$ and standard deviation $a^2\sigma$.
\end{proposition}

\noindent \textbf{Hypothesis Testing.} The estimated scan cycle $T_{ESC}$ vector under normal operation can be represented as $r_k$ for the $k^{th}$ PLC  and for the watermark as $r_k^* = \beta(r_k) + \alpha$, where $\alpha$ is the random delay introduced as a watermark and $\beta$ denotes the scaling of the distribution due to the change in the control logic. Intuitively it means that the estimated scan cycle pattern in $r_k$ is offset with a constant value $\alpha$, an obvious consequence of which is change in the mean of the random variable. Using proposition~\ref{gaussian_dist_proposition}, it can be seen that the resultant vector $(r_k^* = \beta(r_k) + \alpha)$ is a linearly transformed version of $r_k$. The mean for such a change is $\bar{r}_k^* = \alpha + (\beta)\bar{r}_k$ and variance $S_{r^*}^{2} = (\beta)^2 S_{r}^{2}$. For this watermark response protocol, two hypotheses need to be tested, $\mathcal{H}_0$ the \emph{without watermark mode} and $\mathcal{H}_1$ the \emph{with watermark mode} using a K-S test.

\subsection{Kolmogorov-Smirnov~(K-S) Test}

In this study a two-sample K-S test is used. The K-S statistics quantifies a distance between the empirical distribution functions of two samples.
Under a replay attack samples would look like the original trained model without watermark. In the absence of any attack watermark would be preserved and null hypothesis would be rejected.

\begin{figure*}[t]
\centering
\includegraphics[height=8cm,width=18cm,keepaspectratio]{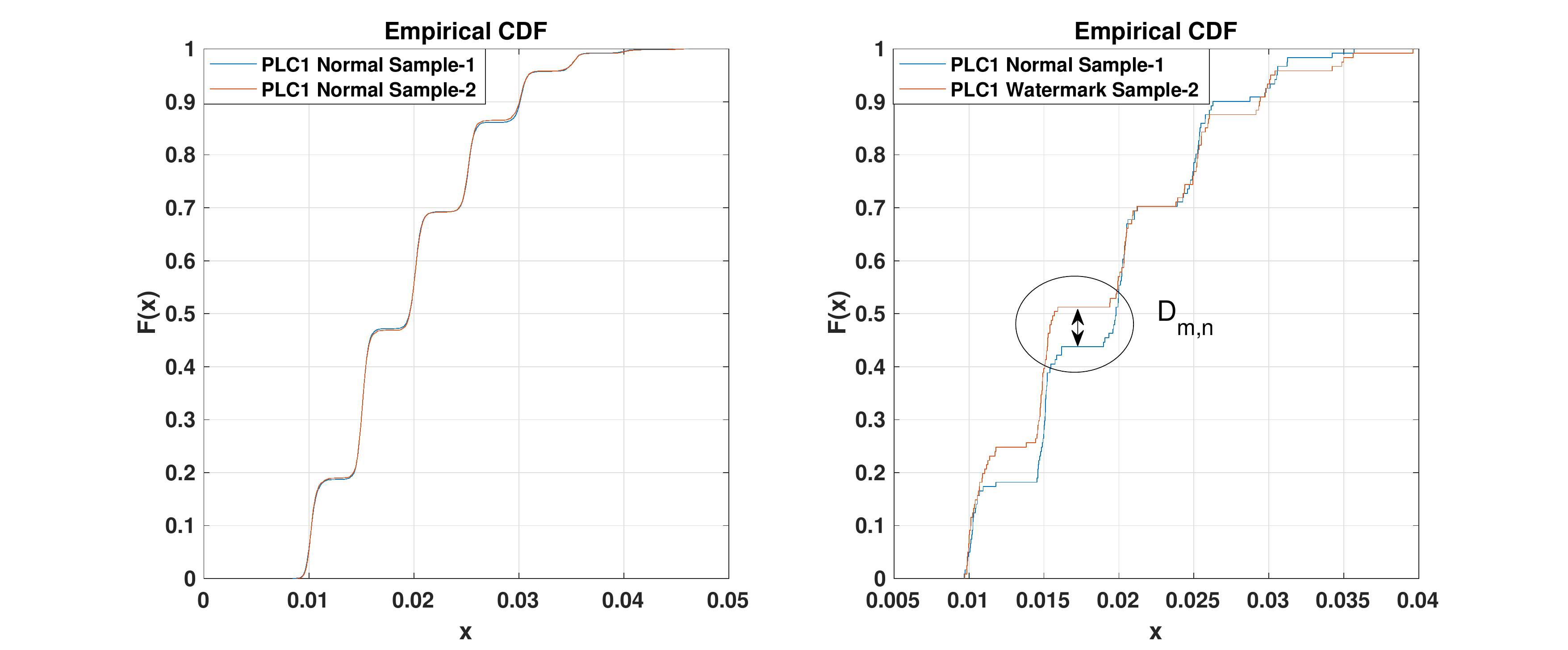}
\caption[PLC1 normal operation empirical CDF]{Gaussian distribution approximation of PLC1 under normal operation for different runs of the experiment. $D_{n,m}$ is the maximum distance between  two distributions.}
\label{with_and_without_watermark_fig}
\vspace{-1em}
\end{figure*}

\subsubsection{K-S test based model training}
In the following, the empirical distributions for the estimated scan cycle time $T_{ESC}$ will be derived and a reference model is obtained without watermarking. There are thousands of samples captured from the PLCs in a matter of a few minutes. If all the captured samples from an experiment are considered it results in a smooth empirical distribution but then the time to make a decision also increases by a few minutes. Therefore, a trade-off between the speed of detection and detection performance is desired.  
Results are  depicted in a tabular form in Table~\ref{kstest_chunksize_vs_accuracy_table} for  PLC~1. For a chunk size of $60$, a TNR of $100\%$ is achieved but, to be little more conservative, a value of $120$ chunk size is chosen in the following analysis.


\begin{table}
\begin{center}

\begin{adjustbox}{max width=0.4\textwidth}
 \begin{tabular}[!htb]{|c | c| c | c | c| c| c | c |} 
 \hline
 Size & 10 &30  &  60  &  120  & 250  &  500  &  1000  \\ 
 \hline
  TNR &  95.57\% & 99.49\% &  100\%  & 100\% & 100\% &100\% & 100\% \\ 
 \hline
 FPR & 4.43\%   & 0.50\%      &   0\% &        0\%    &     0\%   &      0\%  &       0\%  \\  [1ex]
 \hline
 
\end{tabular}
\end{adjustbox}
\end{center}
\vspace{1em}
\caption{K-S test for Normal Data. Chunk size vs classification accuracy for PLC~1.}
\label{kstest_chunksize_vs_accuracy_table}
\vspace{-1.5em}
\end{table}

Two use cases are shown in Figure~\ref{with_and_without_watermark_fig}. For the graph on the left, it can be seen that for two different samples of \sct for PLC1 under  normal operation are very similar, i.e. both the samples are drawn from the same distribution. On the right, one sample is taken from  normal operation and the second sample  from the watermarked \sct of PLC1. It is observed that the distance metric $D_{m,n}$ is greater as compared to the plot on the left hand side, thus these two samples are drawn from  different distributions. This is the key intuition to detect replay attacks in the presence of a watermark signal. 


\begin{table}
\begin{center}

\begin{adjustbox}{max width=0.4\textwidth}
 \begin{tabular}[!htb]{|c | c| c | c | c| } 
 \hline
 WM Type & No-WM &  20~ms delay  &  40~ms delay  &  Rand   \\ 
 \hline
  TNR &  100\% & 91.84\% &  84.62\%   & 98.33\%  \\ 
 \hline
 FPR & 0\%   & 8.16\%      &   15.38\% &        1.67\%      \\  [1ex]
 \hline
 
\end{tabular}
\end{adjustbox}
\end{center}
\vspace{1em}
\caption[K-S test: With and without Watermark]{K-S test with and without adding the watermark.  Accuracy for PLC~1 for a chunk size of $120$ samples is shown.} 
\label{kstest_watermark_Detection_accuracy_table}

\vspace{-1.5em}
\end{table}

\subsubsection{K-S test based model testing}
In this part, testing is done for the dataset from normal operation and for \plcwm. It can be seen from Table~\ref{kstest_watermark_Detection_accuracy_table} that the K-S test produces $100\%$ true negative rate, that is classifying normal data as normal. Second and third columns present results for the case of a static watermark signal. The second column shows the result of injecting  a watermark delay of 20~ms, while the third column shows the results obtained by injecting a watermark delay of 40~ms. In both cases, with  high accuracy, the watermark signals could be classified. The third case is that of a random watermark created using the clock of the PLC. Such a high accuracy of detection motivates the use of K-S test for attack detection. 

\subsection{Watermark Modeling: A Closed Loop Feedback System}
\begin{figure}[t]
\centering
\includegraphics[height=7cm,width=10cm,keepaspectratio]{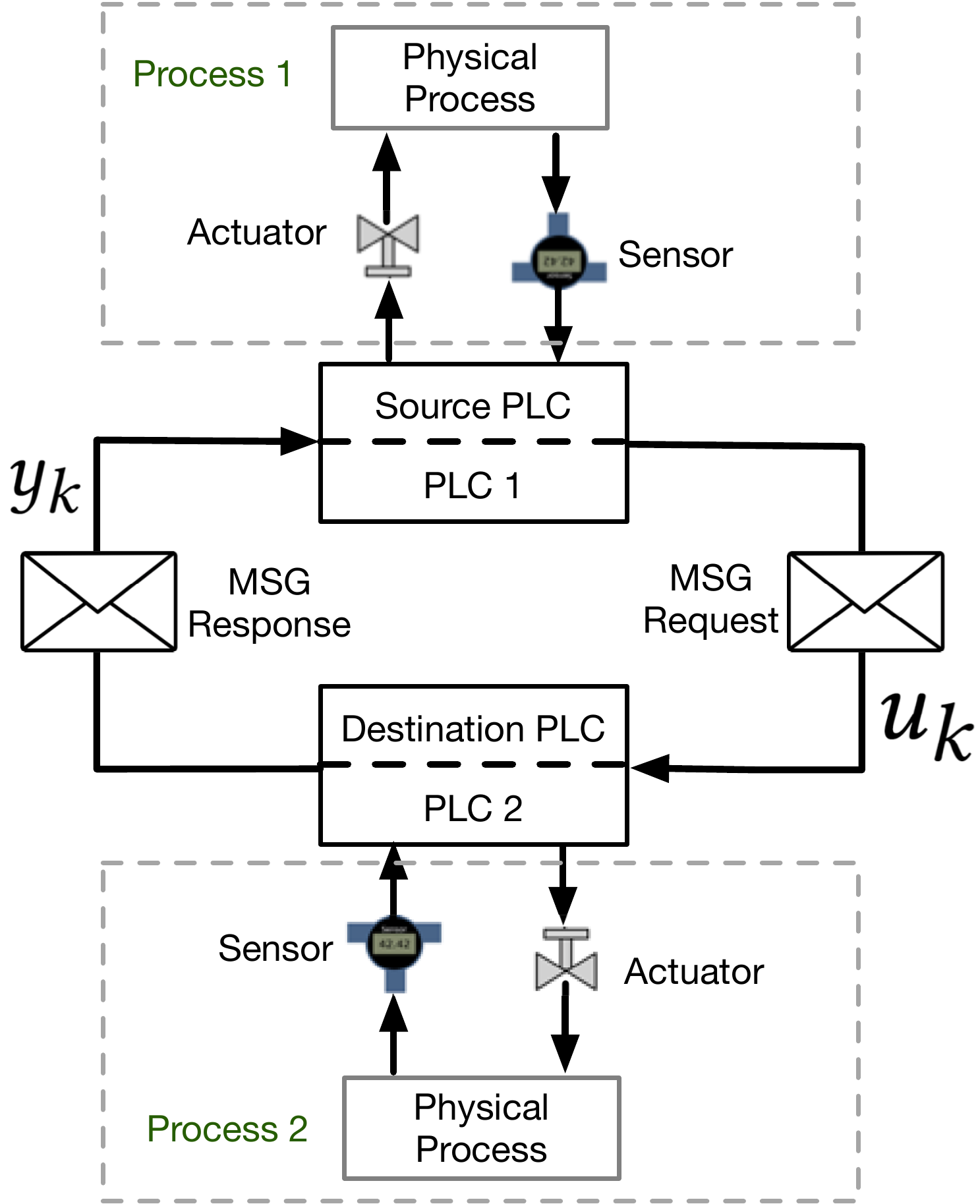}
\caption{Closed loop feedback system model.}
\label{closed_loop_feedback_system_model_fig}
\end{figure}

It is observed that there is a strong relationship between PLC request and response messages as depicted in Figure~\ref{CRP_fig} and Figure~\ref{req_resp_strong_Function_fig}. This process of message exchange can be modeled as a closed loop feedback system from the perspective of control theory. This system model is depicted in Figure~\ref{closed_loop_feedback_system_model_fig}. 

\begin{definition}
The Inter Arrival Time~(IAT)  of MSG requests and responses, respectively,  is defined as the system state referred to as $x_k$, where $k$ is the message number. 
\end{definition}

\begin{definition}
The response MSG time is treated as the output of a system, such as  an output of a sensor, defined as $y_k$.
\end{definition}

\begin{definition}
The dynamics of the scan cycle, PLC hardware and logic complexity govern the dynamics of request messages being sent to other PLCs. These dynamics which are reflected in estimated scan cycle control the timings of receiving the response message. This control action is denoted as $u_k$ where $k$ is the message number.
\end{definition}

\noindent Using subspace system identification methods\,\cite{sensorfaultdetection_systemIdentification2010} the process dynamics can be modeled and represented in a state space form as follows,
\begin{equation}\label{system_state_eq1}
    x_{k+1} = Ax_k + Bu_k + v_k, {\text {and}}
\end{equation}

\begin{equation}\label{output_eq2}
    y_k = Cx_k + \eta_k.
\end{equation}

This is a well-known state space system model which is generally used to model dynamics of a physical process.  In the system of equations~\eqref{system_state_eq1},\eqref{output_eq2}, $y_k$ is the output of the system which is response message timing profile. This output is a function of request messages that  act as a control input. Figure~\ref{req_resp_strong_Function_fig} shows the relationship of response message with the request message timing profile which is driven by the \emph{scan cycle} time. $v_k$ and $\eta_k$ are identical and independently distributed sources of noise due to communication channels. Matrices $A, B$ and $C$ capture the input-output relationship and models the communication among PLCs. The system of equations in \eqref{system_state_eq1} and \eqref{output_eq2} can model the underlying system but it can be subject to powerful attacks for example replay attacks or masquerade attacks. An attacker can learn the system behavior and  replay the collected data while real system state might be reporting different measurements. 

\begin{definition}
{\em PLC Watermarking $\Delta u_k$}:  The output, i.e.  the response MSG time, depends on the request MSG profile. However, request MSG timing profile depends on the scan cycle and other communication overheads. These factors together constitute the  control input $u_k$. A watermark is added to this control input such that the effects of the added watermark are observable on the output of the system, i.e. the response MSG timing profile $y_k$. 
\end{definition}

\begin{proposition}
Replay attack can be detected using the \plcwm technique given the system model of \eqref{system_state_eq1} and \eqref{output_eq2}.
\end{proposition}

\noindent \emph{Proof:} The proposed PLC watermarking technique injects a watermark defined as $\Delta u_k$ in the control signal $u_k$. A replay attack will use the normal data and system model as defined in Eqns.\, \eqref{system_state_eq1} and \eqref{output_eq2}. An attacker, unaware of the watermark, would be exposed  as follows,

\begin{equation}\label{system_state_withwatermark_eq}
    x_{k+1} = Ax_k + B(u_k + \Delta u_k) + v_k, 
\end{equation}

\begin{equation}\label{output_withwatermark_eq}
    y_{k+1} = Cx_{k+1} + \eta_{k+1}.
\end{equation}

Substituting $x_{k+1}$ in the above equation results in,

\begin{equation}\label{output_withwatermark_eq_1}
    y_{k+1} = CAx_{k} +CB u_k + v_{k+1} + \eta_{k+1} + CB \Delta u_k.
\end{equation}

The last term in the above equation is the watermark signal which is  generated randomly using the PLC clock. This watermark signal will expose a replay attack.\hfill $\blacksquare$


\begin{table}
\begin{center}

\begin{adjustbox}{max width=0.7\textwidth}
 \begin{tabular}[!htb]{|c | c| c | c | c| c | c |} 
 \hline
 Attack  / PLC  & PLC~1 & PLC~2  &  PLC~3  &  PLC~4 & PLC~5 & PLC~6   \\ 
 \hline
  Replay: TPR &  95\% & 97.78\% &  82.05\%    & 100\% & 100\% & 98.21\%  \\ 
 
 Replay: FNR   & 5\%  & 2.22\%  &  17.95\%   &  0\% &  0\% &     1.79\% \\  
 \hline
 
   MFDK: TPR &  88.24\%     & 100\%      &  81.25\%   & 100\% &  100\% & 100\% \\ 
 
 MFDK: FNR &       11.76\%   & 0\%      &   18.75\%   &    0\%  & 0\%  & 0\%    \\  [1ex]
 \hline
 
\end{tabular}
\end{adjustbox}
\end{center}
\vspace{1em}
\caption[\plcwm Attack Detection Performance]{K-S test for attack data vs watermark attack detection accuracy for all PLCs in the SWaT testbed for a chunk size of $120$ samples. MFDK: Masquerade Full Distribution Knowledge}\label{kstest_attack_Detection_accuracy_table}
\vspace{-1.5em}
\end{table}

The effects of the watermark in detecting attacks are considered next. Figure~\ref{masquerade_CDF_and_watermark_fig} in Appendix~\ref{metrics_appendix} is an example of powerful masquerade attack. 
In this case the defender is expecting the presence of a watermark in the response received from the other PLC but it did not get that and raised an alarm. 
The same result is shown in   Table~\ref{kstest_attack_Detection_accuracy_table}. The results are  for all  PLCs in the six stages of SWaT. Replay and powerful masquerade attacks are detected with high accuracy using the watermark signals as shown in Figure~\ref{masquerade_CDF_and_watermark_fig}. A  high true positive rate (i.e., attacks declared as attacks), as compared to the results in Table~\ref{mitm_imposter_attack_detection_table}, points to the effectiveness of \plcwm as an authentication technique. In the following, the threat model is further strengthened by assuming that the attacker has the knowledge of the system model and attempts to estimate the watermark signal.

\subsection{System State Estimation}
 
While modeling the system in a state space form, the system states can be estimated using  Kalman filter. This formulation  helps in detecting MiTM attackers which adds same delay in the request~(input) and the response~(output) messages and is also useful in quantifying the contribution of the watermark signal in the response message by normalizing the input and output in terms of the residual. 


\begin{definition}
Let the response MSG timing measurements under a replay attack be  $y_k^a$, control~(request MSG) signal under replay attack  $u_k^a$, and the state estimate  $x_k^a$ at  time step $k$,  $0 < k \leq T$ for an attack time period $T$.
\end{definition}

\begin{proposition}
Given the system of equations for normal system model as \eqref{700}-\eqref{701}, and replay attack defined in Eqns.\,\eqref{800}-\eqref{801} in the Appendix~\ref{system_modeling_appendix}, it can be shown that replay attack would not be detected.
\end{proposition}

\textit{Proof:} The residual vector under an attack is given as,
\begin{equation}
    r_{k+1}^a = y_{k+1}^a - \hat{y}_{k+1}.
\end{equation}
During the replay attack for times $0 < k \leq T$, where T is the time for the readings being replayed, $y^a_{k+1} = y_{k+1}$, resulting in $r^a_{k+1} = r_{k+1}$. Therefore, this results  in no detection and the alarm rate reduces to the false alarm rate of the detector in use. \hfill $\blacksquare$


\begin{figure}[t]
\centering
\includegraphics[height=5.5cm,width=8cm]{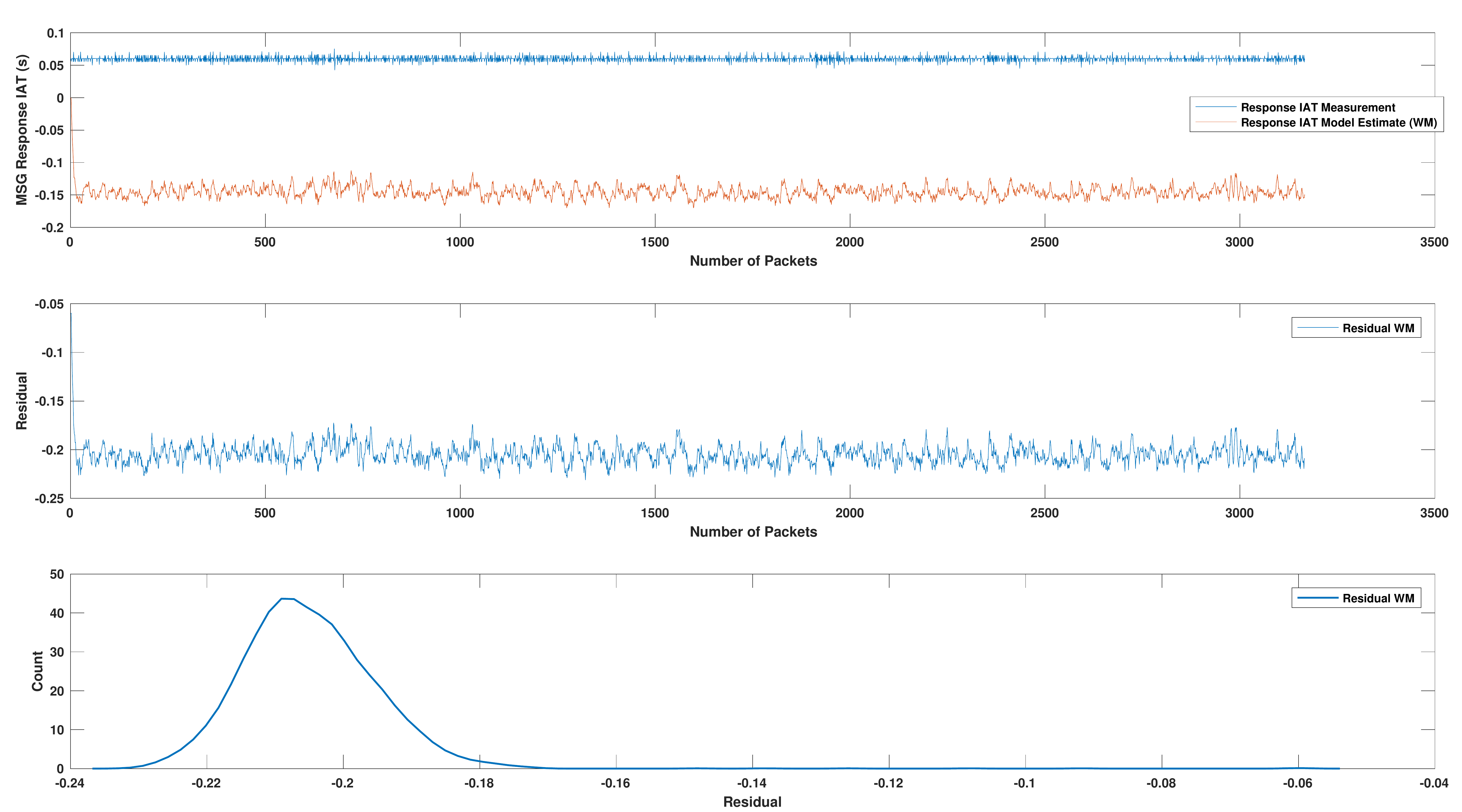}
\caption[PLC1 with a Watermark Signal]{A delay of 40ms is used as a watermark for PLC1.}
\label{modelbased_with_watermark_fig}
\vspace{-1.5em}
\end{figure}


\begin{thm}\label{wm_theorem}
Given the system model   in~\eqref{output_withwatermark_eq_1} and \eqref{output_eq2}, Kalman filter~\eqref{KF_gain} and \eqref{KF_x} as shown in Appendix~\ref{system_modeling_appendix}, and watermarked signal $\Delta u_k$, it can be shown that the residual vector is driven by the watermark signal and can be given as, $r_{k+1} = [CA-CLC](x_k^a - \hat{x}_k^{wm}) + CB(u_k^a - u_k) -CB\Delta u_k + Cv_{k} + \eta_{k+1}$.
\end{thm}


\textit{Proof:} See Appendix~\ref{theorems_appendix}.

The term $-CB(\Delta u_k)$  quantifies the effects of watermark. Based on this it could be investigated  if it is possible to recover the watermark signal from the response message, and  if not then the system is declared as under attack. Figure~\ref{modelbased_with_watermark_fig} shows results from an experiment with a watermarked delay of $40 ms$. In the top pane it can be seen that the watermarked response is different from the estimate of response MSG IAT using the system model developed earlier. In the bottom, a density plot for the residual vector is shown which can be seen as being deviated from the zero mean under the effect of the watermark. 
How much does this distribution deviate depends on the watermark. 


\begin{figure}[t]
\centering
\includegraphics[height=6cm,width=8cm]{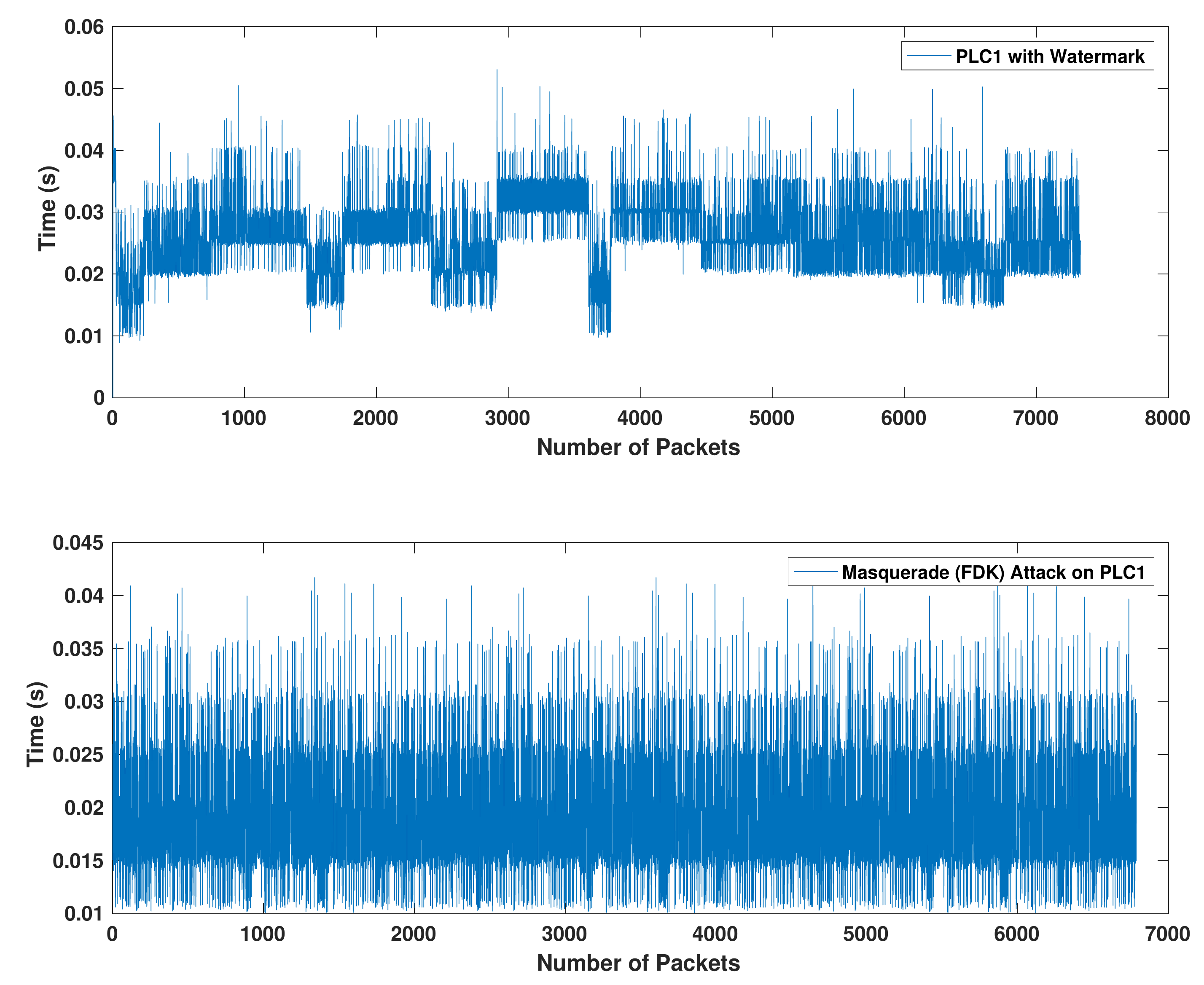}
\caption[Dynamic Watermark and Masquerade Attack]{Masquerade attack and watermark time series. }
\label{masquerade_and_watermark_timeseries_fig}
\vspace{-1em}
\end{figure}

\subsection{A More Powerful Attacker}
In the following analysis we consider the case that attacker is trying to estimate the watermark signal as well. 

\begin{definition}
A watermark signal which is chosen randomly in each iteration of experiment is defined as the dynamic watermark.
\end{definition}


\begin{thm}\label{dynamic_WM_theorem}
An attacker attempting to estimate the watermark signal in the presence of dynamic watermark signal, would be detected because of the hardware delays to switch between the watermarks. System model results in $r_{k+1} = (CA - CLC)(x_k^a - \hat{x}_k^{wm})\\   + CB (u_k^a - u_k)  + CB(\Delta u_k^a - \Delta u_k)  + Cv_k + \eta_{k+1} $, where $\Delta u_k^a$ is attacker's learned watermark signal.
\end{thm}

\textit{Proof} is similar to the proof of theorem~4.1.

In the result from the theorem~\ref{dynamic_WM_theorem} the first three terms are important. The attacker's goal is to make $x_k^a = x_k^{wm}, u_k^a = u_k$ and $\Delta u_k^a = \Delta u_k.$ If this can be achieved, the attacker can hide  amid the use of the watermark signal $\Delta u_k$. The attacker needs  finite  time to obtain the estimate~($\Delta u_k^a$) of the change in the dynamic watermark signal. While doing so, the attacker will be exposed due to the zero transition time required to detect and switch to the new random watermark signal~\cite{shoukry2015}.

\paragraph{Security Argument}
The proposed watermark is an instance of the control logic shown to alter
the  execution time and ultimately the scan cycle time. Therefore, by challenging a PLC at a randomly selected
time $t$ and  duration $\delta$, one would  expect  the estimated scan cycle measurements  to contain the effects of the watermark. If not, then one would suspect that the measurements received, at the network, to be spoofed. An attacker is  aware of such a  \emph{watermark}, but at the same time is spoofing the messages at the network layer, therefore needs  to consistently reflect the watermark profile starting at time $t$ and for duration $\delta$. However, the attacker needs to wait $\Delta$ seconds to recognize that the estimated scan cycle profile has been changed at the beginning and at the end of the watermark. Therefore, the attacker  can at most react consistently with the expectations of the watermark at time ($t + \Delta$) and stop at ($t+\delta+\Delta$). As is shown in the previous sections,  $\Delta$ time units needed to detect if a change in the profile is significant, is approximately 3s in our case study,  and can be leveraged to detect incoherent responses of the attacker to the watermark. 

 A dynamic random watermark signal is shown in Figure~\ref{masquerade_and_watermark_timeseries_fig}. The top pane  shows the response MSG timing profile for a random watermark signal. In the bottom, a masquerader's  strategy is shown for PLC1. From the top pane it can be seen that the watermark is random while it does not affect the system performance as it is a tightly bounded function. 
 A masquerade attacker knows the original system model and inputs/outputs but cannot completely follow the dynamic watermark, thus being exposed.

\section{Related Work}

\noindent\emph{TCP/Network based Fingerprints}: Authors in~\cite{ICS_fingerprinting_2015_networktraffic} used packet size, the frequency of the packets and other network features to create network traffic pattern, which can detect basic attacks but cannot detect a masquerade attack which does not violate these patterns. Network trace and connection patterns are also studied to create fingerprints for the ICS networks~\cite{passive_network_fingerprinting_2016_noDPI,connectionpattern_based_network_fingerprinting}. The methods summarized above borrow the ideas from information technology networks and apply those to ICS networks. However, the challenges in ICS networks are different due to device heterogeneity, proprietary protocols, device computational power and long-standing TCP sessions~\cite{Casali_Dina_feasibility_of_device_fingerprinting}. 

\noindent \emph{PLC Fingerprinting}: The research community has made a few recommendations~\cite{power_fingerprinting_PLC_2014,NIPAD_powerbased_detection_PLC,RF_fingerprinting_PLC_MSthesis,RF_fingerprinting_PLC_malwareDetection}. In~\cite{power_fingerprinting_PLC_2014,NIPAD_powerbased_detection_PLC} a power based profile for PLC logic execution is created. 
In~\cite{RF_fingerprinting_PLC_malwareDetection,RF_fingerprinting_PLC_MSthesis} radio frequency emission by the PLC during the program execution is used to detect malware attacks on the PLCs. 
These related techniques are, 1) invasive, 2) cannot detect cyber and physical attacks on the PLCs and 3) the cost of such systems is prohibitive, for example, the equipment used in \cite{RF_fingerprinting_PLC_MSthesis} costs \$25K at a minimum. Most related work is published recently by Formby et. al~\cite{raheem_formby_scan_cycle_paper_2019} where measuring the program execution times can detect control logic modifications for the PLCs. The measurements are performed within a PLC for a threat model of control logic modifications. Our work is significantly different in that it device a technique to obtain the scan cycle measurements on the network and is able to both authenticate a PLC as well as detect network based intrusions. To the best of our knowledge, the proposed technique in this paper is the first to use unique operational characteristics of  PLCs to passively and non-invasively create network  fingerprints. 
The proposed PLC fingerprinting technique successfully identifies PLCs with an accuracy as high as $99\%$ and attack detection accuracy up to $100\%$.

\section{Conclusions}

A timing based fingerprinting technique is proposed for industrial PLCs. The proposed technique is used to estimate the \emph{scan cycle} time. It is observed that PLCs could be uniquely identified without any modifications to the control logic. It is possible to create unique fingerprints for the same model of PLCs. Powerful attackers with the knowledge of scan cycle time and replay attacks are detected by the proposed \plcwm technique. 

%% file: appendix.tex
\appendix

\section{Theorems and Proofs}
\label{proofs_appendix}
\subsection{Lower bound on eta}\label{lower_bound_eta_appendix}
\textbf{Proposition} \textit{Given the definition of $\eta$ as above, it can be observed that for a fastest PLC it would transmit message each \emph{scan cycle} therefore, required number of scan cycles to transmit a message is at least 1. It can be shown that the $\eta$ is lower bounded as,}
\begin{equation}
\label{eta_eqn_appendix}
   \eta  \geq 1 + \epsilon 
\end{equation}%

\textit{where $\epsilon \in \mathbf{R}^+ $ is a  non-negative real finite number.} 

\emph{\textbf{Proof}:} From~(\ref{eta_eqn_appendix}) it is obvious that on the network we should see the message at least in $1$ \emph{scan cycle} time. This would be the case when \emph{MSG} instruction is executed at each \emph{scan cycle} and all the pre-conditions for the message transmission are fulfilled before the next \emph{scan cycle}. However, let's see the formal proof. We know that estimated \emph{scan cycle} can be expressed as $T_{ESC} = T_{SC} + T_{OverHead}$ and $\eta$ as $\eta = \frac{T_{ESC}}{T_{SC}}$. Then we get,

\begin{equation} 
\label{eta_proof_eq}
    \eta = 1 + \frac{T_{OverHead}}{T_{SC}}
\end{equation}%

\emph{Case 1: if $T_{SC} >> T_{OverHead}$.} In this case the second term~($\epsilon$) in Equation~\ref{eta_proof_eq} comes out to be between 0 and 1. An empirical result is shown in the Table~\ref{responseTime_table}. A controlled experiment is performed by increasing the complexity of the logic to make $T_{SC} >> T_{OverHead}$. In this case, the dominating factor is the \emph{scan cycle} as compared to the overhead. Therefore it can be seen in the Table~\ref{responseTime_table} that all the PLCs take more than 1 but less than 2 scan cycles on average. Since the \emph{scan cycle} time is very high all the pre-conditions for the next packet transmission are fulfilled within a \emph{scan cycle} and packets get transmitted in the next \emph{scan cycle}. $\epsilon$, in this case, represents the network noise. This proves the lower bound.

\emph{Case 2: if $T_{SC} << T_{OverHead}$.} In this case the second term~($\epsilon$) in Equation~\ref{eta_proof_eq} comes out to be greater than 1. An empirical result is shown in the Table~\ref{responseTime_table}. The empirical results reported for this column are from the normal settings of the PLCs in SWaT testbed and represent a normal behavior of the PLCs in a realistic industrial setting. Since each PLC has a different \emph{scan cycle} time and overheads it is not possible to provide a tight upper bound. However, it is shown that a relationship~($\eta$) between the PLC \emph{scan cycle} time~($T_{SC}$) and the estimated \emph{scan cycle} time~($T_{ESC}$) can be established.   \hfill $\blacksquare$

\subsection{Theorems}\label{theorems_appendix}


\textbf{Theorem 4.1} \textit{Given the system model   in~\eqref{output_withwatermark_eq_1} and \eqref{output_eq2}, Kalman filter~\eqref{KF_gain} and \eqref{KF_x}, and watermarked inputs $\Delta u_k$, it can be shown that the residual vector is driven by the watermark signal and can be given as, $r_{k+1} = [CA-CLC](x_k^a - \hat{x}_k^wm) + CB(u_k^a - u_k) -CB\Delta u_k + Cv_{k} + \eta_{k+1}$.}\label{wm_theorem_app1}



\textit{\textbf{Proof}:} In the system model of eq.~\eqref{system_state_eq1}, attacker has access to the normal timing measurements and control~(request) signals and can replay those. Assuming that an adversary has access to the system model, Kalman filter gain and the parameters of the detector. During the replay attack using its knowledge an attacker tries to replay the data resulting in a system state described as,

\begin{equation}
    {x}_{k+1}^a = Ax_k^a +Bu_k^a + v_k 
\end{equation}
 and attacker's spoofed sensor measurements as,
 \begin{equation}
     {y}_{k}^a = C{x}_{k}^a + \eta_k
 \end{equation}
 
 \begin{equation}
     y_{k+1}^a = C[Ax_k^a + Bu_k^a + v_k] + \eta_{k+1}
 \end{equation}
 
 \begin{equation}
           y_{k+1}^a = CAx_k^a + CBu_k^a + Cv_k + \eta_{k+1}
 \end{equation}

 However, using a watermarking signal in the control input $u_k$, the defender's state estimate becomes,
 \begin{equation}
     \hat{x}_{k+1}^{wm} = A\hat{x}_k^{wm} + Bu_k +B\Delta u_k + L(y_k^a - C\hat{x}_k^{wm}),
 \end{equation}
where $\Delta u_k$ is the watermark signal. 

\begin{equation}
    \hat{y}_{k+1}^{wm} = C\hat{x}_{k+1}^{wm}
\end{equation}

\begin{equation}
    \hat{y}_{k+1}^{wm} = C[A\hat{x}_k^{wm} + Bu_k +B\Delta u_k + L(y_k^a - C\hat{x}_k^{wm})]
\end{equation}

\begin{equation}
    \hat{y}_{k+1}^{wm} = CA\hat{x}_k^{wm} + CBu_k +B\Delta u_k + CL(Cx_k^a - C\hat{x}_k^{wm})]
\end{equation}

\begin{equation}
    \hat{y}_{k+1}^{wm} = CA\hat{x}_k^{wm} + CBu_k + B\Delta u_k + CLCx_k^a - CLC\hat{x}_k^{wm}]
\end{equation}

The residual vector in the presence of a replay attack is given as,

\begin{equation}
    r_{k+1} = y_{k+1}^a - \hat{y}_{k+1}^{wm}
\end{equation}

\begin{equation}
    \begin{split}
    r_{k+1} = CAx_k^a + CBu_k^a + Cv_k + \eta_{k+1} - CA\hat{x}_k^{wm} - CBu_k \\ - CB\Delta u_k - CLCx_k^a + CLC\hat{x}_k^{wm}
    \end{split}
\end{equation}

\begin{equation}
\begin{split}
    r_{k+1} = (CA - CLC)x_k^a - (CA-CLC)\hat{x}_k^{wm} + CB(u_k^a - u_k)\\ -CB(\Delta u_k) + Cv_k + \eta_{k+1}
\end{split}
\end{equation}

\begin{equation}
    \begin{split}
    r_{k+1} = (CA - CLC)(x_k^a - \hat{x}_k^{wm}) + CB(u_k^a - u_k) + Cv_k \\ + \eta_{k+1} - CB(\Delta u_k) 
    \end{split}
\end{equation}

The last term above is the watermark signal and the  first term is the error. For a stable system the spectral radius of ($CA-CLC < 1$) and the error converges to zero\,\cite{Astrom}. \hfill $\blacksquare$

\subsection{K-S Test}
\label{ks_test_appendix}

The K-S statistics quantifies a distance between the empirical distribution functions of two samples. 

\begin{definition}
\emph{Empirical Distribution Function:} For $n$ independent and identically distributed ordered observations of a random variable $\mathbf{X}$, an empirical distribution function $F_n$  is defined as, 
\begin{equation}
    F_n(x) = \frac{1}{n}\sum_{i-1}^n I_{[-\infty,x]}(\mathbf{X_{i}}),
\end{equation}
where $I_{[-\infty,x]}(\mathbf{X_{i}})$ is the indicator function which is $1$  if $\mathbf{X_i} \leq x$, else it is equal to 0. 
\end{definition}

For a given Cumulative Distribution Function~(CDF) $F(x)$, the K-S statistic is given as,
\begin{equation}
    D_n =  \sup_{x \in \mathbb{R}} | F_n(x) - F(x) |,
\end{equation}
where $\sup$ is the supremum of the set of distances. For the two-sample K-S test the statistic can be defined as,
\begin{equation}
        D_{n,m} =  \sup_{x \in \mathbb{R}} | F_n(x) - G_m(x) |,
\end{equation}
where $F_n(x)$ and $G_m(x)$ are the empirical distribution functions of the first and second sample, respectively. $D_{n,m}$ is the maximum of the set of distances between the two distributions. For a large sample size the null hypothesis is rejected for the confidence level $\alpha$ if,
\begin{equation}
    D_{n,m} > c(\alpha) \sqrt{\frac{n+m}{n*m}},
\end{equation}
where $n,m$ are, respectively,  the sizes of the first and second samples. The value of $c(\alpha)$ can be obtained from the look up tables for different values of $\alpha$, or can be calculated as follows,
\begin{equation}
    c(\alpha) = \sqrt{-\frac{1}{2}ln(\alpha)}.
\end{equation}

\section{System Modeling}
\label{system_modeling_appendix}

\subsection{Kalman filter}\label{kf_appendix}

Given the system model in Eqns.~\eqref{system_state_eq1} and \eqref{output_eq2}, the state of a system can be estimated based on the available output $y_k$, using  a linear Kalman filter with the following structure,
\begin{equation}
\hat{x}_{k+1} = A \hat{x}_k + B u_k + L_k \big( \bar{y}_k - C\hat{x}_k   \big),  \label{estimation_eq}
\end{equation}
with estimated state $\hat{x}_k \in \mathbb{R}^n$, $\hat{x}_1 = E[x(t_1)]$, where $E[\hspace{.5mm}\cdot\hspace{.5mm}]$ denotes the expectation, and gain matrix $L_k \in \mathbb{R}^{n \times m}$. The  estimation error is defined as  $e_k:= x_k - \hat{x}_k$. In Kalman filter matrix $L_k$ is designed to minimize the covariance matrix $P_k:= E[e_ke_k^T]$ in the absence of attacks.

Eq.~\eqref{estimation_eq} is an overview of the system model where the Kalman filter is being used for estimation. The estimator makes an estimate at each time step based on the previous readings up to $x_{k-1}$ and the sensor reading $y_k$. the estimator gives $\hat{x}_k$ as an estimate of state variable $x_k$. Thus, an error can be defined as,
 \begin{equation}\label{error}
e_k=\hat{x}_k - x_k,
\end{equation} 
where $\hat{x}_{k | j}$ denotes the optimal estimate for $x_k$ given the measurements $y_{1},...,y_{j}$. Let $P_k$ denote the error covariance, $Cov(e_k)=E[(x_k - \hat{x}_k)(x_k - \hat{x}_k)^{T}]$, and $\hat{P}_{k | j}$  the estimate of $P_k$ given $y_{1},...,y_{j}$. Prediction equation for state variable using Kalman filter can be written as,
 \begin{equation}\label{KF_x}
\hat{x}_{k+1 | k}=A\hat{x}_{k | k} 
\end{equation} 
 \begin{equation}\label{KF_P}
P_{k + 1| k}=AP_{k | k}A^{T} + Q,
\end{equation} 
where $\hat{x}_{k | k}$ is the estimate at time step $k$ using measurements up to time $k$ and $\hat{x}_{k+1 | k}$ is $(k+1)^{th}$ is the prediction based on previous $k$ measurements. Similarly, $P_{k | k}$ is the error covariance estimate until time step $k$. $Q$ is the process noise covariance matrix. The next step in Kalman filter estimation is time update step using Kalman gain $L_k$. 
 
 \begin{equation}\label{KF_gain}
L_k=P_{k | k - 1}C^{T}(CP_{k | k - 1}C^{T} + R)^{-1}
\end{equation}

 \begin{equation}\label{KF_update_x}
\bar{x}_{k + 1 | k}=\hat{x}_{k + 1 | k} + L_k(y_k - C\hat{x}_{k + 1 | k})
\end{equation} 

 \begin{equation}\label{KF_P_update}
\bar{P}_{k + 1 | k}=(I - L_kC)\hat{P}_{k + 1 | k},
\end{equation} 
where $\bar{x}_{k + 1 | k}$ and $\bar{P}_{k + 1 | k}$, are the updates for the $k + 1$ time step using measurements $y_{i}$ from the $i^{th}$ sensor and Kalman gain $L_k$. $R$ is the measurement noise covariance matrix. The   initial state can be selected  as $x_0=x_{0}$ with $P_0=E[(\hat{x}_{0} - x_{0})(\hat{x}_{0} - x_{0})^{T}]$. Kalman gain $L_k$ is updated at each time step but after a few iterations it converges and operates in a steady state. Kalman filter is an iterative estimator and $\hat{x}_{k | k}$ in equation~\ref{KF_x} comes from $\bar{x}_{k - 1 | k}$ in equation~\ref{KF_update_x}. It is assumed that the system is in a steady state before attacks are launched.  Kalman filter gain is represented by $L$ in steady state.  


\begin{figure}[t]
\centering
\includegraphics[height=7cm,width=9cm]{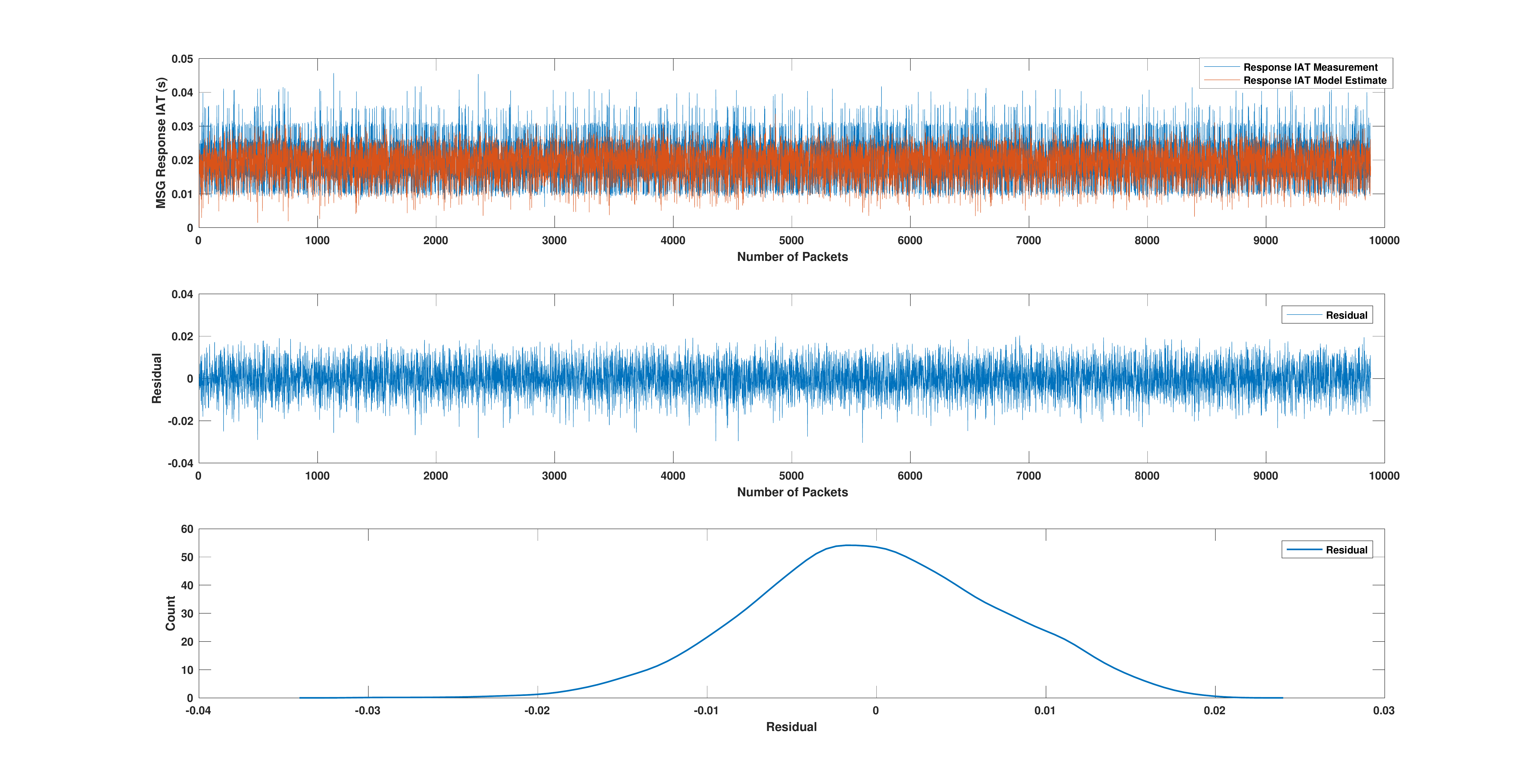}
\caption[PLC1 MSG Response without Watermark]{PLC1 MSG response without the watermark signal. Under the derived system model, the residual has a Gaussian distribution with zero mean. A watermark signal can be quantified and then extracted from the response message using this normal profile.}
\label{modelbased_without_watermark_fig}
\end{figure}


\begin{figure}[htp]
\centering
\includegraphics[scale=0.3]{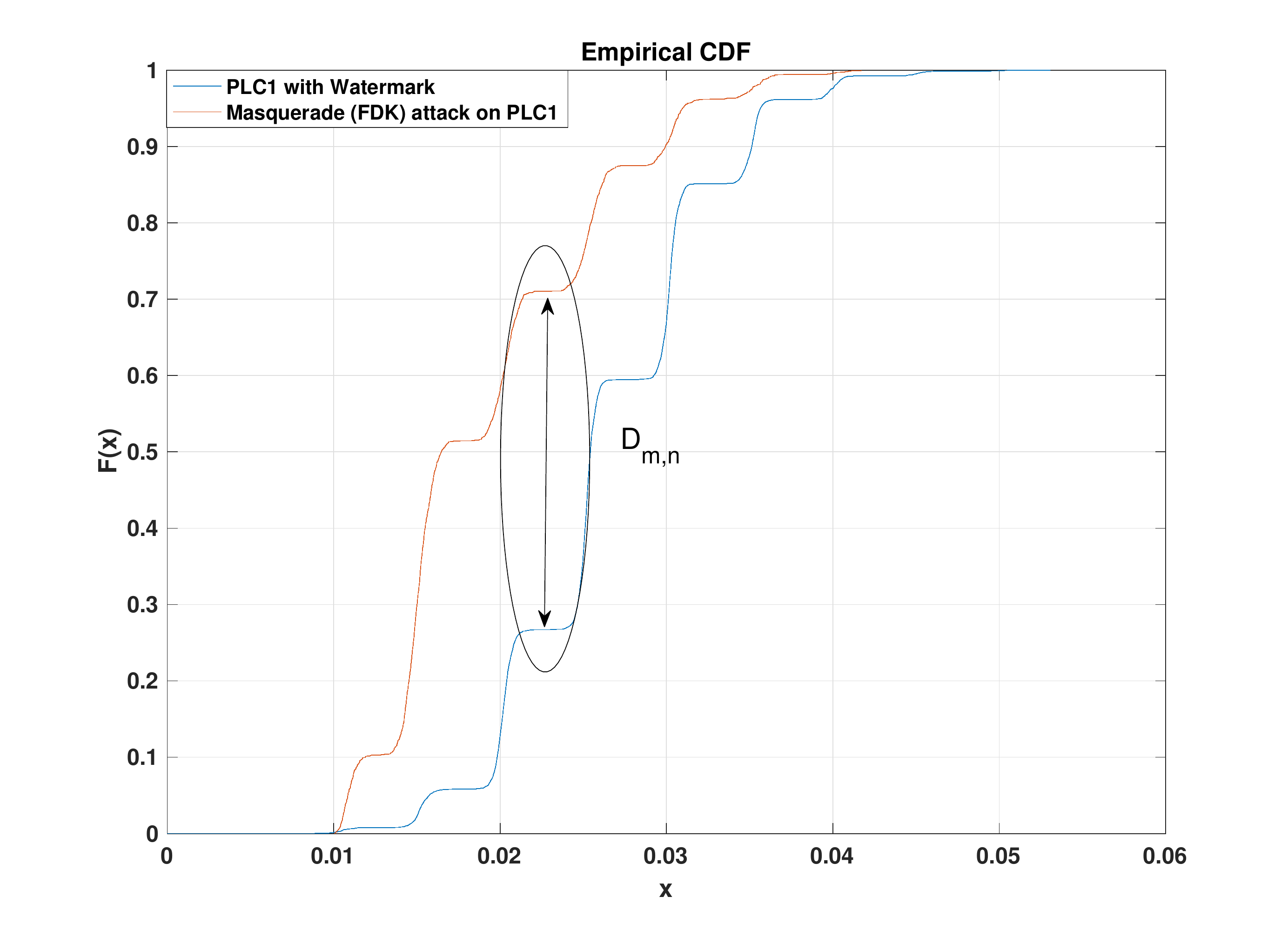}
\caption{Masquerade attack on PLC1 and watermark. These cumulative distributions show that due to the watermark a masquerader exposes itself via a K-S test.}
\label{masquerade_CDF_and_watermark_fig}
\end{figure}

\begin{figure}[htp]
\centering
\includegraphics[height=7cm,width=9cm,keepaspectratio]{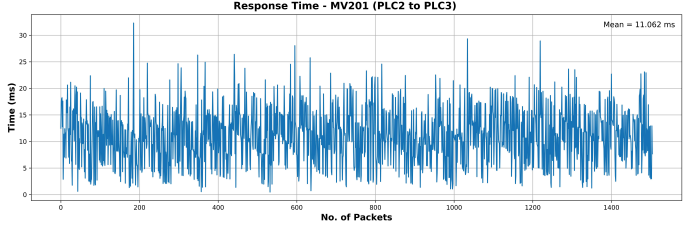}
\caption{Response time for a message instruction in PLC3 in SWaT testbed.}
\label{responseTime_fig}
\end{figure}

\subsection{Residuals and hypothesis testing}
The estimated system state is compared  with timing measurements $\bar{y}_k$ which may have the presence of an attacker. The difference between the two should stay within a certain threshold under normal operation, otherwise, an alarm is triggered. The  residual random sequence $r_k, k\in \mathbb{N}$ is defined as,
\begin{align}
r_k := {y}_k - C\hat{x}_k = Ce_k + \eta_k + \delta_k. \label{25}
\end{align}
If there are no attacks, the mean of the residual is
\begin{equation}
E[r_{k+1}] = CE[e_{k+1}] + E[\eta_{k+1}] = \bar{r}_{m \times 1} = 0_{m \times 1}.  \label{27} 
\end{equation}
where $\bar{r}_{m \times 1}$ denotes an $m \times 1$ matrix composed of mean of residuals under normal operation, and the co-variance is given by
\begin{align}
\Sigma := E[r_{k+1}r_{k+1}^T] &= CPC^T + R_2 .\label{28}
\end{align}
For this residual, hypothesis testing is done, $\mathcal{H}_0$ the \emph{normal mode}, i.e., no attacks, and $\mathcal{H}_1$ the \emph{faulty mode}, i.e., with attacks. 
The residuals are obtained using this data together with the state estimates. Thus, the two hypotheses can be stated as follows, 
\begin{center}
\begin{tabular}{ c c c }
 $\mathcal{H}_0: \left\{
\begin{array}{ll}
E[r_k] = \bar{r}_{m \times 1},  \label{29_5} \\[.5mm]
E[r_kr_k^T] = \Sigma, 
\end{array} \text{or}
\right.$ & $\mathcal{H}_1: \left\{
\begin{array}{ll}
E[r_k] \neq \bar{r}_{m \times 1},\   \label{30_5} \\[.5mm]
E[r_kr_k^T] \neq \Sigma.
\end{array}
\right.$
\end{tabular}
\end{center}

\noindent The system model under the normal operation and in the presence of a replay attack can be given as,

\noindent\rule{\hsize}{1pt}\vspace{.5mm}
\textbf{Normal Operation:} 
\begin{equation}
\left\{
\begin{array}{ll}
X_{k+1} = Ax_k + Bu_k + v_k,  \label{700}   \\
y_k = Cx_k + \eta_k ,\text{ \ \ \ \ \ \  \hspace{4.6125mm} } {\text {system model}}.
\end{array}
\right.
\end{equation}

\begin{equation}
\left\{
\begin{array}{ll}
\hat{x}_{k+1} = A\hat{x}_k + Bu_k + L_k(y_k - C\hat{x}_k), \label{701}   \\
\hat{y}_{k} = C\hat{x}_{k}, \text{ \ \ \ \ \ \  \hspace{6.6125mm} state estimation}.
\end{array}
\right.
\end{equation}


\noindent\rule{\hsize}{1pt}\\[.5mm]

\noindent\rule{\hsize}{1pt}\vspace{.5mm}
\textbf{Replay Attack:} 
\begin{equation}
\left\{
\begin{array}{ll}
X_{k+1}^a = Ax_k^a + Bu_k^a + v_k,  \label{800}   \\
y_k^a = Cx_k^a + \eta_k ,\text{ \ \ \ \ \ \  \hspace{4.6125mm} } {\text {system model}}.
\end{array}
\right.
\end{equation}

\begin{equation}
\left\{
\begin{array}{ll}
\hat{x}_{k+1}^a = A\hat{x}_k^a + Bu_k^a + L_k(y_k^a - C\hat{x}_k^a), \label{801}   \\
\hat{y}_{k}^a = C\hat{x}_{k}^a,  \ \ \ \ \ \  \hspace{6.6125mm}{\text{ state estimation}}.
\end{array}
\right.
\end{equation}


\noindent\rule{\hsize}{1pt}\\[.5mm]

\section{Performance Metrics and Supporting Figures}
\label{metrics_appendix}

Each PLC is assigned a unique ID and multi-class classification is applied to identify it among all the PLCs.   Identification accuracy is used as a performance metric. Let $c$ denote the total number of classes,  $TP_i$ the true positive for class $c_i$ when it is rightly classified, $FN_i$  the false negative defined as the incorrectly rejected,  $FP_i$ the false positive as incorrectly accepted, and $TN_i$  the true negative as the number of correctly rejected PLCs. The overall accuracy ($acc$)  is defined as follows. 

\begin{equation}
acc = \frac{\sum_{i=1}^{c} TP_i+\sum_{i=1}^{c}TN_i}{\sum_{i=1}^{c}TP_i+\sum_{i=1}^{c}TN_i+\sum_{i=1}^{c}FP_i+\sum_{i=1}^{c}FN_i}.
\end{equation}

\noindent The True Positive Rate (TPR) and False Positive Rate (FPR) are  defined as follows.

\begin{equation}\label{tpr}
\mbox{True Positive Rate (TPR)} = \frac{TP}{TP + FN},
\end{equation}
\begin{equation}\label{fpr}
\mbox{False Positive Rate (FPR)} = \frac{FP}{FP + TN}.
\end{equation}

\begin{equation}\label{fnr}
\mbox{False Negative Rate (FNR)} = 1- TPR.
\end{equation}
\begin{equation}\label{tnr}
\mbox{True Negative Rate (TNR)} = 1- FPR.
\end{equation}


\begin{table}
\begin{center}
\caption{List of features used. Information Value~(IV) helps to choose features based on the values that contribute significantly to the classification accuracy and those not bringing any unique information for classification would be dropped.}
\label{features}
\begin{adjustbox}{max width=0.4\textwidth}
 \begin{tabular}[!htb]{|l | l| l |} 
 \hline
 {\bf Feature} & {\bf Description} & {\bf IV} \\ 
 \hline
 Mean & $\bar{x} = \frac{1}{N} \sum_{i=1}^{N} x_i$ & 0.5 \\ 
 \hline
 Std-Dev & $\sigma = \sqrt[]{\frac{1}{N-1}\sum_{i=1}^{N}(x_i - \bar{x}_i)^2}$ & 0.5 \\
 \hline
 Mean Avg. Dev & $D_{\bar{x}} = \frac{1}{N}\sum_{i=1}^{N} |x_i - \bar{x}|$  & 0.5 \\
 \hline
 Skewness & $\gamma = \frac{1}{N}\sum_{i=1}^{N}(\frac{x_i - \bar{x}}{\sigma})^3$ & 0.041 \\
 \hline
 Kurtosis & $\beta = \frac{1}{N} \sum_{i=1}^{N}(\frac{x_i - \bar{x}}{\sigma})^4 - 3$ & 0.063 \\
 \hline
 Spec. Std-Dev & $\sigma_s = \sqrt[]{\frac{\sum_{i=1}^{N}(y_f(i)^2)*y_m(i)}{\sum_{i=1}^{N}y_m(i)}}$  & 0.5\\
 \hline
 Spec. Centroid & $C_s = \frac{\sum_{i=1}^{N}(y_f(i))*y_m(i)}{\sum_{i=1}^{N}y_m(i)}$  & 0.5\\
 \hline
 DC Component & $y_m(0)$ &  0.6\\ 
 \hline
 Spectral Crest & $SC_s = \frac{(Max_{i=1 to N}(y_m(i)))}{C_s}$ & 0.42\\ [1ex] 
 \hline
 Smoothness & $S_s = \sum_{i=2}^{N-1} |20.log(y_m(i)) - $ & 0.5\\
 & $\frac{20(log(y_m(i-1)) + log(y_m(i)) + log(y_m(i+1)))}{3}|$ & \\
 \hline
 
 Spectral Flatness & $F_s = \frac{(\prod_{i=1}^N y_m(i))^{1/N}}{\frac{\sum_{i=1}^{N} y_m(i)}{N}}$ & 0.5\\
 \hline
 Spectral Skewness & $\gamma_s = \frac{\sum_{i=1}^{N}(y_m(i) - C_s)^3 * y_m(i)}{\sigma_s^3}$ & 0.5\\
 \hline
 Spectral Kurtosis & $\beta_s = \frac{\sum_{i=1}^{N}(y_m(i) - C_s)^4 * y_m(i)}{\sigma_s^4 - 3}$ & 0.5\\
 \hline
 
 \multicolumn{2}{p{0.45\textwidth}}{ $x$: time domain data from a sensor for $N$ elements in the data chunk;  $y_f$:  bin frequencies; $y_m$: magnitude of the frequency coefficients.} 
\end{tabular}
\end{adjustbox}
\end{center}
\end{table}